\documentclass[3p,times]{elsarticle}

\usepackage{ecrc}

\volume{00}

\firstpage{1}

\journalname{Nuclear Physics A}

\runauth{}

\jid{procs}

\jnltitlelogo{Nuclear Physics A}

\CopyrightLine{2023}{Published by Elsevier Ltd.}

\usepackage{amssymb}

\usepackage[figuresright]{rotating}
\usepackage{bbding}
\usepackage{amsmath}
\usepackage{amssymb}
\usepackage{latexsym}
\usepackage{graphics}      
\usepackage{slashed}
\usepackage{color}
\usepackage{comment}
\usepackage{cancel}
\usepackage{subfig}
\usepackage{tikz}
\usetikzlibrary{shapes,arrows}
\usetikzlibrary{intersections}
\usepackage[colorlinks=true,citecolor=cyan,urlcolor=blue,bookmarks=true,bookmarksopen=true,bookmarksnumbered=true,bookmarksopenlevel=3]{hyperref}
\pdfoutput=1
\definecolor{Orange}{cmyk}{0.,0.5,0.5, 0.}

\newcommand{\tK}{\tilde K}
\usepackage{setspace}

\biboptions{sort&compress}
\begin{document}

\begin{frontmatter}
	
	
	
	\dochead{}
	
	\title{Probing DDM and ML quantum concepts in shape phase transitions of $\gamma $-unstable nuclei}
	\author[label1]{S. Ait El Korchi}
	\ead{salwa.aitelkorchi@ced.uca.ma}
	\author[label1]{M. Chabab}
	\ead{mchabab@uca.ac.ma}
	\author[label1]{A. El Batoul\fnref{cor1}}
	\fntext[cor1]{Corresponding author}
	\ead{abdelwahed.elbatoul@uca.ac.ma}
	\author[label1,label2]{A. Lahbas}
	\ead{a.lahbas@um5r.ac.ma}
	\author[label1]{M. Oulne}
	\ead{oulne@uca.ac.ma}
	\address[label1]{High Energy Physics and Astrophysics Laboratory, Department of Physics, Faculty of Sciences Semlalia, Cadi Ayyad University P.O.B 2390, Marrakesh 40000, Morocco.}
	\address[label2]{ESMaR, Faculty of Sciences, University Mohammed V in Rabat, Rabat, Morocco.}
	\begin{abstract}
		In a recent paper (S. Ait El Korchi et al. 2020 EPL 132 52001), we explored, inside the context of Critical Point Symmetries (CPSs) X(3) and Z(4), a correlation between two exceedingly known quantum concepts, the Minimal Length (ML) and the Deformation-Dependent Mass (DDM), that are commonly  applied in various areas of physics. Such a correlation has been strongly identified in transition nuclei by calculating some physical observables of that  quantum system, like as energy spectra, moments of inertia and transition probabilities. In this paper we extend that study to E(5) dynamical symmetry corresponding to the shape phase transition U(5)$\leftrightarrow$O(6). The experimental realization of the models  was found to occur in some nuclei, using the existing phenomenological potentials : Infinite Square Well, Davidson and Kratzer, whose models fits provide the best agreement. Importantly the calculations performed in this work using  these potentials, corroborate the fact that the revealed correlation between both quantum concepts is not destructively affected by the presence of other model’s parameters and hence its existence is independent of the form or type of the used potential. Undoubtedly, the present work will open the way for more investigations of this correlation in the limits of other critical points symmetries in nuclear shape phase transitions which play today a major role in nuclear structure research from theoretical as well as experimental point of view.
	\end{abstract}
	
	\begin{keyword}
		Bohr–Mottelson model\sep Critical point symmetries\sep Shape phase transitions\sep Collective models.
		\PACS 21.60.Ev\sep21.60.Fw\sep 21.10.Re.
		
	\end{keyword}
	
\end{frontmatter}


\section{Introduction}
To understand and be able to correctly describe the properties of the atomic nucleus, we need a consistent theory of its structure. One phenomenological example of this is the Bohr-Mottelson model (BMM) \cite{b1} and its extensions. In the previous two decades, the amount of attention paid to its solutions has increased tremendously, and this has occurred for a variety of reasons. One of these is represented by the appearance of its cornerstone solutions, called  E(5) \cite{b2}, X(5)\cite{b3}, Y(5)\cite{b4} and Z(5)\cite{b5}. In the same perspective, research involving the Bohr-Mottelson Hamiltonian concentrated more on understanding shape phase transitions and associated critical point symmetries (CPSs), but in the last two years, a new orientation of its use for the sake of shape coexistence, fluctuations  and mixing phenomena has been proposed \cite{b6,b7,b8,b9,b10,b11}. Furthermore, critical point symmetries have typically paved the way for the development of new models by combining various phenomenological potentials, resulting in new precisely or quasi-exactly separable models that can describe nuclei that are close to or far from existing critical point symmetries. 
\par The BMM\cite{b1}, including its theoretical underpinnings, provided us with a geometrical alternative theory of nuclear collective excitations, as opposed to the algebraic models\cite{b12,b13}. In essence, the geometric Bohr Mottelson model depicts the collective excitations of nuclei in terms of liquid drop surface oscillations.
In the large majority of cases, the resolution of its Hamiltonian necessitates the employment of a potential $V(\beta,\gamma)$ that is completely reliant on the parameters $\beta$ and $\gamma$, where $\beta$ indicates ellipsoidal deformation and $\gamma$ is the measure of axial asymmetry. Based on scientific studies, this realistic theoretical model was able to depict successfully the low energy collective states and the electromagnetic transitions of a large number of even-even nuclei\cite{b13-a,b13-a1}. 
\par In the framework of BMM, the nuclear surface of the deformed nucleus is an ellipsoid arbitrarily oriented in space and described, for our purposes, by a second-order deformation such as \cite{b14}:
\begin{equation}
	R(\theta,\phi)=R_0\left[ 1+ \sum^{+2}_{\mu=-2}\alpha_{2,\mu}Y_{2,\mu}(\theta,\phi) \right],
	\label{eq1}
\end{equation}
where $\alpha_{2,\mu}$ are tensors describing the deformations of the nucleus. They are expressed in terms of  the radius of the spherical nucleus $R_0$ and the spherical harmonics $Y_{\lambda,\mu}$ as follows 
\begin{equation}
	\alpha_{2,\mu}=(-1)^{\mu}\alpha^*_{2,-\mu}=\frac{1}{R_0}\int R(\theta,\phi)Y_{2,\mu}(\theta,\phi)d\Omega.
	\label{eq2}
\end{equation}
However, it is indespensable to transform into a coordinate system which is "fixed" in the oscillating body. The collective coordinates in the body-fixed system are then connected to the space-fixed system by the following transformation \cite{b14}:
\begin{equation}
	a_{2,\mu}=\sum_{\mu}\alpha_{2,\mu}\mathcal{D}_{\mu,\nu}\left(\theta_i\right),
	\label{eq3}
\end{equation}
where the $\mathcal{D}_{\mu,\nu}\left(\theta_i\right)$ are the Wigner-D functions for the spherical harmonics of second order and 
the triad of Eulerian angles ($\theta$, $\varphi$, $\psi$) describing the relative orientation of the axes is here symbolized by $\theta_i$. The ellipsoid about the principal axes in the proper coordinate frame, imposes that $a_{2,1}=a_{2,-1}=0$ and $a_{2,2}=a_{2,-2}$. Thus the five variables $a_{\lambda,\mu}$ are replaced by the three Eulerian angles $\theta_i$ and  the two real internal coordinates $a_{2,0}$ and $a_{2,2}$. Again, in place of $a_{2,0}$ and $a_{2,2}$ it is beneficial to introduce the variables  $\beta$ and $\gamma$ employing both closed relations : $a_{2,0}=\beta\cos\gamma$ and $a_{2,2}=a_{2,-2}=(\beta/\sqrt{2})\sin\gamma$, which are obviously respected the conventions of D.L. Hill and J. A. Wheeler \cite{b15}.
In view of these conventions, the collective Bohr Hamiltonian for quadrupole shapes in the five-dimensional form is written as \cite{b1,b14}:
\begin{equation}
	H=T+V(\beta, \gamma),
	\label{eq4}
\end{equation}
with,
\begin{align}
	T=\frac{-\hbar^2}{2B_m}\Bigg[ \frac{1}{\beta^4}\frac{\partial}{\partial\beta}\beta^4\frac{\partial}{\partial \beta}+\frac{1}{\beta^2 sin3\gamma}\frac{\partial}{\partial\gamma}sin3\gamma \frac{\partial}{\partial \gamma} -\frac{1}{4\beta^2}\sum_{k}\frac{Q^2_{k}}{sin^2(\gamma-\frac{2}{3}\pi k)}\Bigg],
	\label{eq4m}
\end{align}
where $B_m$ is the mass parameter, while $Q_{k}$ represents the angular momentum in the variables $\theta_i$. 
According to the selected form for the potential $V(\beta,\gamma)$, there are three basic situations in which one can accomplish a perfect separation of the variables, one of them should be considered: potentials independent \cite{b16} of the collective variable $\gamma$, called $\gamma$-unstable potentials, appropriate for describing vibrational and near vibrational nuclei. 
On the other hand, in terms of the form, it might be argued that the appropiate differential equation to be solved is, in some ways, nothing more than the Schr\"{o}dinger equation, and that the record of examples known in the quantum mechanical setting also applies to the current situation. This is somewhat correct, however there are a number of significant distinctions that must be highlighted and for which the collective Hamiltonian requires particular consideration: because the Bohr equation is expressed in terms of two variables (rather than five) as previously indicated, its natural space is 5-dimensional rather than 3-dimensional, and this has an impact on not only the asymptotic behavior and boundary conditions, but also the group structure that we associate with the Bohr Hamiltonian. 
\par Although shape/phase transitions result in abrupt changes in nuclear characteristics, so critical point symmetries aiming at defining the transition point must match to reliable spectra and B(E2) transition rates predictions. 
In the present work we study a $\gamma$-unstable  Bohr hamiltonian, in the critical point symmetry E(5) which is designed to describe the shape phase transition from spherical U(5) to $\gamma$-unstable O(6) shapes. We will however consider three phenomenological potentials: the infinite square well, Davidson and Kratzer in the presence of two formalisms: the deformation dependent mass (DDM) and the minimal length (ML), while pointing out that these last are largely used and tested in Bohr's Hamiltonian solution fields \cite{b17,b18,b19,b20,b21,b22,b23,b24,b25,b26,b27,b28,b28a,b29,b30}. Moreover, in a recent work \cite{b29} we have revealed the existence of a strong correlation between both formalisms that we will also check in the present study. The solutions obtained will be dubbed E(5)-ML-IFSW,  E(5)-ML-D,  E(5)-ML-K for the ML formalism within the infinite square well, Davidson and Kratzer potentials respectively, and in the same way we called E(5)-DDM-IFSW,  E(5)-DDM-D and  E(5)-DDM-K for the DDM formalism.
The expressions for the energy spectrum as well as the wave functions are obtained by means of the asymptotic iteration method (AIM) \cite{b31} in the DDM case, and by using conjointly AIM and  quantum perturbation method (QPM) \cite{b32} for the ML.
\par The present paper is outlined as follows: after the introduction which provides the background of the  critical point symmetries (CPSs) as well as  the physics content of Bohr's Hamiltonian, we present in Section II the Bohr Hamiltonian with deformation-dependent effective mass formalism, while in Section III we give a brief description of it but this time with minimal Length. The basic equations of E(5)  involving deformation-dependent mass term with different phenomenological potentials and their solutions are  given in Section IV.  The models E(5) including the concept of minimal length with the same potentials and their solutions are also exposed in Section V. The B(E2) transition probabilities are considered in Section VI. Finally, Section VII is devoted to the numerical computations and discussion of theoretical results  with experimental data for energy spectra, B(E2) transition probabilities and moment of inertia, while Section VIII contains the conclusions.
\section{Bohr Hamiltonian with deformation-dependent effective mass} 
Deformation-dependent effective mass \cite{b33} is widely used in quantum physics and it is equivalent to a deformation of the canonical commutation relations:
\begin{equation}
	[x_i, x_j] = 0,\ 
	[x_i, p_j] =  {\rm i} \hbar \delta_{i,j},\ 
	[p_i, p_j] =  0 \; ,  
	\label{eq5}
\end{equation}
where $i=1,2,3$. By replacing the momentum components
$p_i = - {\rm i} \hbar \nabla_i= - {\rm i} \hbar \partial/\partial x_i$  by some deformed hermitian operators:
\begin{equation}
	\pi_i = \sqrt{f(x)}\, p_i \sqrt{f(x)} \; ,  
	\label{eq6}
\end{equation}
where the positive real deforming function $f(x)$ depends on the coordinates $ x=(x_{1} ,x_{2},x_{3})$, both last commutators in Eq.\eqref{eq5}  transform into :
\begin{eqnarray}
	[x_i, \pi_j]  = {\rm i} \hbar f(x) \delta_{i,j}, \
	[\pi_i, \pi_j]  = {\rm i} \hbar [f_j(x) \pi_i-f_i(x) \pi_j],
	\label{eq7}
\end{eqnarray}
with $f_i(x) \equiv \nabla_i f(x)$.
Bohr equation with a mass depending on the deformation coordinate $\beta$ describing the quadrupole vibrations and rotations in a unified manner, can be easily derived  by using classical expression of Hamiltonian and Pauli-Podolsky prescription \cite{b34} for the canonical quantization in curvilinear coordinates,
\begin{equation}
	(\nabla \Phi)^i = g^{ij} {\partial \Phi \over \partial x^j}, \qquad 
	\nabla^2 \Phi = {1\over \sqrt{g}} \partial_i \sqrt{g} g^{ij} \partial_j \Phi,
	\label{eq8}
\end{equation} 
equipped with the collective Bohr-Wheeler coordinates $\beta$,$\gamma$ and the three Euler angles $\Omega$ which are connected with the
rotational angles $\omega_k$ by a linear transformation.
Because the deformation function $f$ is solely dependent on the radial coordinate $\beta$, only the $\beta$ component of the arising equation will be altered. The final result reads\cite{b18}  

\begin{align}\label{eq9}
	\Big[ 
	-{1\over 2} {\sqrt{f}\over \beta^4} {\partial \over \partial \beta} 
	\beta^4 f {\partial \over \partial \beta} \sqrt{f}
	-{f^2 \over 2 \beta^2 \sin 3\gamma} {\partial \over \partial \gamma} 
	\sin 3\gamma {\partial \over \partial \gamma}  &+ {f^2\over 8 \beta^2} 
	\sum_{k=1,2,3} {Q_k^2 \over \sin^2\left(\gamma -{2\over 3} \pi k \right)}+ v_{eff}(\beta,\gamma) \Big] \Psi = \epsilon \Psi,  
\end{align}
where reduced energies $\epsilon = B_m E/\hbar^2$ and reduced potentials
$v(\beta,\gamma)= B_m V/\hbar^2$ have been used, with
\begin{eqnarray}
	\label{eq10}
	v_{eff}(\beta,\gamma)= v(\beta,\gamma)+ {1\over 4 } (1-\delta-\lambda) f \nabla^2 f +{1\over 2} \left({1\over 2} -\delta\right) \left( {1\over 2} -\lambda\right)
	(\nabla f)^2. 
\end{eqnarray}
The model's free parameters $\lambda$ and  $\delta$  are introduced originally by O. von Roos\cite{b34add} in order to overcome the ambiguity problem in ordering of the kinetic energy operator, namely to identify effective potential\eqref{eq10} resulting from all possible choices of ordering which unambiguously own the same solutions.
\section{Bohr Hamiltonian with minimal Length } 
The concept of ML can be implemented on the study of the physical systems by taking into account the deformed canonical commutation relation
\cite{b35,b36,b37}:
\begin{equation}
	[\hat{X}, \hat{P}]=i h\left(1+\alpha^{2} \hat{P}^{2}\right),
	\label{eq11}
\end{equation}
where $\alpha$ representing the ML parameter is a very small positive parameter. This commutation relation yields the following uncertainty relation: 
\begin{equation}
	(\Delta \hat{X})(\Delta \hat{P}) \geq \frac{\hbar}{2}\left(1+\alpha(\Delta \hat{P})^{2}+\hat{\tau}\right), 
	\hat{\tau} =\alpha\langle\widehat{P}\rangle^{2}.
	\label{eq12}
\end{equation}
The above conception has been employed in numerous quantum physical theories that have suggested the existence of a ML having magnitudes ranging from the Planck constant ($10^{-35}m$) \cite{b38,b39} to the order $10^{-15}m$  \cite{b40,b41,b42}. In this light, the ML concept has been proposed in the pioneering study\cite{b25}. Therefore, the collective quadrupole Hamiltonian of Bohr-Mottelson with minimal length is as follows: 
\begin{equation}
	H=-\frac{\hbar^2}{2B_m}\Delta+\frac{\alpha \hbar^4}{B_m}\Delta^2+V(\beta,\gamma),
	\label{eq13}
\end{equation}
with $\Delta$ formulated in the following way :
\begin{equation}
	\Delta={\hbar^2\over \sqrt{g}} \partial_i \sqrt{g} g^{ij} \partial_j,
	\label{eq14}
\end{equation}
her $g$ is the determinant of the symmetric matrix $g_{i j}$ having the form:
\begin{eqnarray}\label{eq15}
	(g_{ij})= \left( \begin{matrix} g_{11} & g_{12} & g_{13} & 0 & 0 \cr
		g_{21} & g_{22} &   0    & 0 & 0 \cr
		g_{31} &   0    & g_{33} & 0 & 0 \cr
		0    &   0    &   0    & g_{44} & 0 \cr 
		0    &   0    &   0    & 0  & g_{55} \cr \end{matrix}\right),  
\end{eqnarray}
with \cite{b14}
\begin{align}\label{eq16}
	& g_{11}= {{\cal J}_1 \over B_m} \sin^2\Theta \cos^2\psi +  {{\cal J}_2\over B} \sin^2\Theta \sin^2\psi
	+ {{\cal J}_3\over B_m} \cos^2\Theta, \nonumber \\
	& g_{12}= {1\over B_m} ({\cal J}_2-{\cal J}_1) \sin\Theta \sin\psi \cos\psi ,\nonumber  \\
	& g_{13}= { {\cal J}_3 \over B_m} \cos\Theta, \nonumber \\
	& g_{22}= {{\cal J}_1 \over B_m} \sin^2\psi + {{\cal J}_2 \over B_m} \cos
	^2\psi,  \\
	& g_{33}= {{\cal J}_3 \over B_m}, \nonumber \\
	& g_{44}=1 ,\nonumber \\
	& g_{55}=\beta^2,\nonumber
\end{align}   
where the moments of inertia are :
\begin{equation}
	{\cal J}_k = 4 B_m \beta^2 \sin^2\left( \gamma - k {2\pi \over 3}  \right). 
	\label{eq17}
\end{equation}
The Laplacian operator \eqref{eq14} in terms of collective variables $(\beta,\gamma)$ can be written as : 

\begin{eqnarray}
	\Delta=\frac{1}{\beta^{4}}\frac{\partial}{\partial\beta}\beta^{4}\frac{\partial}{\partial\beta}+\frac{1}{\beta^2 sin3\gamma}\frac{\partial}{\partial\gamma}sin3\gamma \frac{\partial}{\partial \gamma}-\frac{1}{4\beta^2}\sum^{3}_{k=1}\frac{Q_{k}^{2}}{sin^{2}(\gamma-\frac{2\pi}{3}k)} ,
	\label{eq17a}
\end{eqnarray}

At the first order on $\alpha$, the collective Schr\"{o}dinger equation corresponding to the Hamiltonian\eqref{eq13} then reads as 
\begin{equation}
	\left[ -\frac{\hbar^{2}}{2B_{m}}\Delta+\frac{\alpha \hbar^{4}}{B_{m}}\Delta^{2}+V\left(\beta,\gamma\right)-E_{n,L}\right] \Psi(\beta,\gamma,\Omega) =0.
	\label{eq18}
\end{equation}
Using the definition of reduced energies and those for potentials, also the following transformation 
\begin{equation}
	\Psi(\beta,\gamma,\theta_{i})=(1+2\alpha\hbar^{2}\Delta)\zeta(\beta,\gamma,\theta_{i}),
	\label{eq19}
\end{equation}
the equation \eqref{eq18} in $\gamma$-unstable case is simplified to
\begin{equation}
	\left[(1+4B_{m}\alpha(E-v(\beta))) \Delta+(\epsilon_{n,L}-v(\beta))\right] \zeta(\beta,\gamma,\Omega)=0,
	\label{eq20}
\end{equation}
and  the separation of variables is assumed	by taking  the following  wave function 
\begin{equation} \zeta(\beta,\gamma,\theta_{i})=\xi(\beta) \, \Phi(\gamma,\theta_{i}).
	\label{eq21}
\end{equation}
The radial part $ \xi(\beta) $ obeys to
\begin{equation}
	\left[ \frac{1}{\beta^{4}}\frac{\partial}{\partial\beta}\beta^{4}\frac{\partial}{\partial\beta}+\frac{\Lambda}{\beta^2}+\frac{2B_{m}}{\hbar^2}\bar{K}(E,\beta)\right] \xi(\beta) =0,
	\label{eq22}
\end{equation}
with $ \bar{K}(E,\beta) $ is given by
\begin{equation}
	\bar{K}(E,\beta)=\frac{E-v(\beta)}{1+4B_{m}\alpha(E-v(\beta))}.
	\label{eq23}
\end{equation}	
\section{E(5)-DDM models}
\subsection{E(5)-DDM with Infinite Square Well potential}
It is important to mention here that the signatures of the E(5) symmetry are derived theoretically from parameter-free calculations using the five-dimensional infinite well potential in the $\beta$ shape variable, simulating a second-order phase transition between the spherical and $\gamma$-unstable phases.Therefore, since our goal is to study the E(5) model in the presence of the two concepts of DDM and ML, the first potential model that of course comes to our mind and which we should study is the model potential known as IFSW which has an anharmonic behaviour and its form is defined by :
\begin{eqnarray}
	u(\beta) = \left\{ \begin{array}{lcl} 0, && \rm{if}\ 0\leq\beta\leq\beta_{\omega} \\
		\infty, &&  \rm{if}\ \beta > \beta_{\omega} \end{array} \right. ,
	\label{eq24}
\end{eqnarray}
where $\beta_{\omega}$ specifies the width of the well.
For this potential, the suitable deformation function $f(\beta)$ is given by\cite{b29} : 
\begin{equation}
	f(\beta)=\frac{1}{\beta^{a}}, \quad a<<1.
	\label{eq25}
\end{equation}
The eigenvalues and eigenfunctions of the Hamiltonian of equation  \eqref{eq9} with IFSW potential are obtained by considering  the following  transformation $\xi(\beta)=\beta^{(a-\frac{3}{2})}{F}(\beta)$,
\begin{equation}
	E_{s,L} = \frac{\hbar^2}{2B_0}\bar{k}_{s,\tilde{\eta}}^2,\ \ \bar{k}_{s,\tilde{\eta}}= \frac{(1+a)}{\beta_{\omega}^{(1+a)}}\chi_{s,\tilde{\eta}} \ ,
	\label{eq26}
\end{equation}
and 
\begin{eqnarray}
	\xi_{sL}(\beta)= N_{s,L}\,\beta^{a-3/2}
	J_{u}\left(\frac{\bar{k}_{s,u}}{a+1}\beta^{a+1}\right),\quad u=\frac{\tilde{\eta}}{a+1},
	\label{eq27}
\end{eqnarray}
where $N_{s,L}$ is a normalization constant of the wave function. \\
The energy spectrum is characterized by the principal quantum number $s$ together with the total angular momentum $L$. The parameter $\tilde{\eta}$ is given by 
\begin{equation}
	\tilde{\eta}^2=\Lambda+a^{2}-3a+\frac{9}{4},
	\label{eq28}
\end{equation}
where  $\chi_{s,u}$ is the $s$-th zero of the Bessel function of the first kind $J_{\eta}(\bar{k}_{s,u}\beta_{\omega})$.
\subsection{E(5)-DDM with Davidson potential}
In this part we are going to consider the Davidson potential  \cite{b43} with two parameters,
\begin{equation}
	u(\beta) = c\beta^2+\frac{\beta_0}{\beta^2},
	\label{eq29}
\end{equation}
this form is a special case of the Davidson potential, where $c$ and $\beta_0$ are two free scaling parameters, and $\beta_m=\left(\frac{\beta_0}{c}\right)^{1/4}$ represents the position of the minimum of the potential. For $\beta_0=0$ and $c=1$, the original solution of Bohr \cite{b1}, without any formalism, which corresponds to a 5-dimensional (5-D) harmonic oscillator characterized by the symmetry U(5) $\supset$ SO(5) $\supset$ SO(3) $\supset$ SO(2) \cite{b44}, is obtained. 
Using the AIM \cite{b31} we get the generalized formula of the  energy spectrum
\begin{eqnarray}
	E_{n,\tau}=\frac{\hbar^2}{2B_m}\Big[k_0+\frac{a}{2}(3+2\sigma_{1}+2\sigma_{2}+\sigma_{1}\sigma_{2})+2a(2+\sigma_{1}+\sigma_{2})n+4an^2 \Big],\ \ 
	\label{eq30}
\end{eqnarray}
here $n$ is the principal quantum number of $\beta$ vibrations and
$k_0$,$k_2$, $k_{-2}$,$\sigma_{1}$ and $\sigma_{2}$ are given by 
\begin{align}
	&k_{0\  }=2 a\Big[6 +\Lambda\Big],   \nonumber\\
	&k_{2\  }=a^2\Big[  \Lambda   +10   \Big]+2c,   \nonumber\\
	&k_{-2}= \Lambda+2+2\beta_0^4, \\
	&\sigma_{1}=\sqrt{1+4k_{-2}},\nonumber \\ 
	&\sigma_{2}=\sqrt{1+4\frac{k_2}{a^2}},\nonumber 
	\label{eq31}
\end{align}
while $\Lambda$ is the eigenvalue of the second-order Casimir operator of
SO(5) given by 
\begin{equation}
	\label{eq32}
	\Lambda=\tau(\tau+3),
\end{equation}
with $\tau$ is the seniority quantum number \cite{b45}, characterizing
the irreducible representations of SO(5). The values of angular momentum $L$ occurring for each $\tau$ are provided by a well known algorithm and are listed in \cite{b16,b46}. Within the ground state band (gsb) one has $L=2\tau$. It should be noted that the used appropriate deformation function  for Davidson potential has the form \cite{b18,b21} :
\begin{equation}
	f(\beta)=1+a\beta^{2}, \quad a<<1.
	\label{eq33}
\end{equation}
This choice is taken so as to arrive at an exact solution to the problem.
\subsection{E(5)-DDM with Kratzer potential}
Here we are going to consider the Kratzer potential \cite{b48}
\begin{equation}
	u(\beta) = \frac{-1}{\beta}+\frac{\tilde B }{\beta^2},
	\label{eq34}
\end{equation}
for this potential the deformation function is given by \cite{b19}
\begin{equation}
	f(\beta)=1+a\beta, \quad a<<1 .
	\label{eq35}
\end{equation}
So, in $\gamma$-unstable nuclei, the energy spectrum of Kratzer reads \cite{b19} 
\begin{equation}
	E_{n,\tau} =-{\hbar^2\over 2B_m \left(n+{1\over 2} +  \sqrt{ \left(\tau+{3\over 2} \right)^2+2\tilde B} \right)^2},  
	\label{eq36}
\end{equation}
and the corresponding excited state wave function having the following form :
\begin{equation}\label{eq37}
	\xi_n(\beta) = N_n \beta^{- \mu_n} f^{\frac{1}{2} \left(\mu_n + \frac{\tK}{\mu_n} - 1\right)} 
	P_n^{\left(\mu_n - \frac{\tK}{\mu_n}, \mu_n + \frac{\tK}{\mu_n}\right)}\left(\frac{2+a\beta}{a\beta}\right),
\end{equation}
where $\tK = k_{-2} - \frac{k_{-1}}{a}$ and $\mu_n=\mu-n$, and by $P_n^{(\alpha,\beta)}(t)$ the Jacobi polynomials \cite{b49} are denoted, while the normalization coefficient $N_n$ is taken from Ref \cite{b19}.
\section{E(5)-ML models}
\subsection{E(5)-ML with IFSW}
In the case of IFSW potential \eqref{eq24}, the collective equation \eqref{eq22} of states becomes
\begin{equation}
	\Biggl[\frac{d^2}{d\beta^2} +
	\frac{4}{\beta}\frac{d}{d\beta}
	+ \Biggl(2\bar{K}+\frac{\Lambda}{\beta^2}\Biggr)\Biggr]
	\xi(\beta) = 0 .
	\label{eq38}
\end{equation}
This equation is solved by using a wave function in the form $ \xi(\beta)=\beta^{-\frac{3}{2}}f(\beta)$,
then it becomes
\begin{equation}
	\Biggl[\frac{d^2}{d\beta^2} +
	\frac{1}{\beta}\frac{d}{d\beta}
	+ \Biggl(\bar{k}^2-\frac{\eta^2}{\beta^2}\Biggr)\Biggr]
	f(\beta) = 0,
	\label{eq39}
\end{equation}
with
\begin{equation}
	\eta^2=\frac{9}{4}+\Lambda \ \text{and} \ \bar{k}^2=\frac{2B_{m}}{\hbar^2}\bar{K}.
	\label{eq40}
\end{equation}
The eigenfunction of the differential equation \eqref{eq38}, which is finite at $ \beta=0 $, is given by
\begin{equation}
	\xi(\beta)=N.\beta^{-\frac{3}{2}}.J_{\eta}(\bar{k}\beta),
	\label{eq41}
\end{equation}
and the corresponding energy septra are therefore obtained:
\begin{equation}
	E_{s,L}=\frac{\hbar^2}{2B_{m}}\frac{\left({	\frac{\chi_{s,\eta}}{\beta_{w}}}\right)^{2}}{1-2\alpha \left(\frac{\chi_{s,\eta}}{\beta_{w}}\right)^{2}},
	\label{eq42}
\end{equation}
where $\chi_{s,\eta}$ is the s-th zero of the Bessel function of the first kind $ J_{\eta}(\bar{k}\beta) $.
\subsection{E(5)-ML with Davidson potential}
In this paragraph we have jointly employed AIM and quantum perturbation theory as methods of resolution in order to solve the collective equation \eqref{eq18}, since this equation is not soluble for the Davidson potential \eqref{eq29}. In this context, we emphasize that we have already obtained good results using these methods in previous works \cite{b28,b29,b30}. An analytical formula for the energies of the ground, $\beta$ and $\gamma$ bands  was then derived within the framework of E(5) symmetry with a Davidson harmonic oscillator potential in $\beta$ shape variable. The energy spectrum appears such as this: 
\begin{equation}
	E_{n,L}=E_{n,L}^{(0)}+\Delta E_{n,L},
	\label{eq43}
\end{equation}
with $ E_{n,L}^{(0)} $ is the unperturbed energy 
\begin{equation}
	E^{0}_{n,L}=\sqrt{\frac{\hbar^{2}}{2B_{m}}}\rho(4\mu+2+4n),
	\label{eq44}
\end{equation} 
where
\begin{equation}
	\mu=\frac{1}{2}+\frac{1}{2}\sqrt{4\Lambda+8\beta_{0}^{2}+9}  
	\,\,\, ,\rho=\sqrt{\frac{a}{2}}.
	\label{eq45}
\end{equation}
It is also possible to get unperturbed wave functions: 
\begin{align} \label{eq46}
	\xi(\beta)=N_{n,L} \beta^{\mu-4}e^{-\rho\beta^{2}} {}_{1}F_{1}(-n;\mu+\frac{1}{2};2\rho\beta^{2})
	=N_{n,L} \beta^{\mu-4}e^{-\rho\beta^{2}}\mathcal{L}_{n}^{(\mu-\frac{1}{2})}(2\rho\beta^{2}),
\end{align}
where $\mathcal{L}_{n}^{(\mu-\frac{1}{2})}(2\rho\beta^{2})$ are the Laguerre polynomials and $N_{n,L}$ the normalization constant.
The quantity $\Delta E_{n,L}$ is the correction to the energy spectrum  induced by the minimal length  using the perturbation theory at the first order. It is given by :
\begin{align} \label{eq47}
	\Delta E_{n,L}=\frac{\alpha \hbar^{4}}{B_{m}}<\psi^{0}(\beta,\theta_{i})\mid\Delta^{2}\mid \psi^{0}(\beta,\theta_{i})>=4\alpha B_{m}\big[ ( E^{0}_{n,L})^2+2c\beta_{0}-2  E^{0}_{n,L}(c \overline{\beta^{2}}+\beta_{0} \overline{\beta^{-2}})+(c^2 \overline{\beta^{4}}+\beta_{0}^2 \overline{\beta^{-4}})\big],
\end{align}
where $ \psi^{0}(\beta,\theta_{i}) $ are the eigenfunctions, solutions to the Schr$\ddot{o} $dinger equation for $\alpha=0$.
The quantities $\overline{\beta^{i}} (i=2,-2,4,-4)$  are expressed as follows :
\begin{eqnarray}
	\left\lbrace
	\begin{aligned}
		&\overline{\beta^{2}}=\frac{4n+1+2\mu}{4\rho},  \\
		&\overline{\beta^{-2}}=\frac{4\rho}{2\mu-1},  \\
		&\overline{\beta^{4}}=\frac{4\mu^{2}+24n^{2}+8\mu+36n+19}{16\rho^{2}},  \\
		&\overline{\beta^{-4}}=\frac{16\rho^{2}(4n+1+2\mu)}{(2\mu+1)(4\mu^{2}-8\mu+3)},  
	\end{aligned}
	\right. 
	\label{eq48}
\end{eqnarray}
The corrected wave function  can be also determined by means of a quantum perturbation method as  
\begin{eqnarray}
	\label{eq49}
	F^{Corr}_{n}(\beta)=F_{n}(\beta)+\sum_{k\neq n}\left[\frac{\int_{0}^{\infty}\beta^2F_{k}(\beta)\vartheta(n,\alpha,c,\beta_{0},E_{n,L}^{(0)})F_{n}(\beta)d\beta}{E_{n,L}^{(0)}-E_{k,L}^{(0)}}\right]F_{k}(\beta), \ \ \ \ 
\end{eqnarray}
with,
\begin{eqnarray}
	\label{eq50}
	\vartheta(n,\alpha,c,\beta_{0},E_{n,L}^{(0)})= 4B_m\alpha\bigg[ ( E^{0}_{n,L})^2+2c\beta_{0}-2  E^{0}_{n,L}(c \overline{\beta^{2}}+\beta_{0} \overline{\beta^{-2}})
	+(c^2 \overline{\beta^{4}}+\beta_{0}^2 \overline{\beta^{-4}})\bigg].
\end{eqnarray}
\subsection{E(5)-ML with Kratzer potential}
In the case of ML with Kratzer's potential in the form \eqref{eq34}, the collective equation\eqref{eq18} has no analytical solution, so to solve it we proceed as in the Davidson potential. For the unperturbed energy one obtains :
\begin{equation}
	E^{0}_{n,\tau}=\frac{\hbar^{2}}{2B_{m}}\left[\frac{1}{2(n+\mu)}\right]^{2} \; ,
	\label{eq51}
\end{equation}
with
\begin{equation}
	\mu=\frac{1}{2}+\sqrt{\tilde{B}+\left(\tau+\frac{3}{2}\right)^{2}},
	\label{eq52}
\end{equation}
and the unperturbed wave function is given by
\begin{eqnarray}
	F(\beta)=&N_{n,L}.\beta^{\mu-2}e^{-\rho\beta}\mathcal{L}_{n}^{(2\mu-1)}(2\rho\beta)
	\label{eq53},
\end{eqnarray}
where $ \mathcal{L}_{n}^{(2\mu-1)} $ denotes the associated Laguerre polynomials and $  N_{n,L}$ is a normalization constant given by
\begin{eqnarray}
	N_{n,L}=\left[\frac{(2\rho)^{\left(2\mu+1\right)}\Gamma(n+1)}{\Gamma(n+2\mu)(2n+2\mu)}\right]^{\frac{1}{2}}.
	\label{eq54}
\end{eqnarray}
The tolal energy spectrum in this case has the form :
\begin{eqnarray}
	E_{n,\tau}=E^{0}_{n,\tau}+\Delta E_{n,\tau},
	\label{eq55}
\end{eqnarray}
where $ \Delta E $ is given by
\begin{align}
	\Delta E_{n,\tau}
	=4\alpha B_{m}\bigg[ ( E^{0}_{n,L})^2+&2E^{0}_{n,L}\overline{\beta^{-1}}+\overline{\beta^{-2}}(1-2AE^{0}_{n,L})-2A\overline{\beta^{-3}}+A^2\overline{\beta^{-4}}\bigg],
	\label{eq56}
\end{align}
where the quantities $\overline{\beta^{i}} (i=-1,-2,-3,-4)$ are expressed as follows
\begin{eqnarray}
	\left\lbrace
	\begin{aligned}
		&\overline{\beta^{-1}}=\frac{1}{2}.\frac{k}{n+\mu},  \\
		&\overline{\beta^{-2}}=\frac{1}{2}.\frac{k^2}{(2\mu+1)(\mu+n)},  \\
		&\overline{\beta^{-3}}=\frac{1}{4}.\frac{k^3}{\mu(\mu-1)(2\mu-1)},  \\
		&\overline{\beta^{-4}}=\frac{1}{4}.\frac{k^4(\mu+3n)(2\mu+1)+3n(n-1)}{\mu(4\mu^{2}-1)(\mu-1)(2\mu-3)(\mu+n)},  \\
	\end{aligned}
	\right. 
	\label{eq57}
\end{eqnarray}
The corrected wave function can be calculated using Eq.\eqref{eq49} with
\begin{align}
	\vartheta(n,\alpha,c,\beta_{0},E_{n,L}^{(0)})=4\alpha B_{m}\bigg[ ( E^{0}_{n,L})^2+2E^{0}_{n,L}\overline{\beta^{-1}}+\overline{\beta^{-2}}(1-2AE^{0}_{n,L})-2A\overline{\beta^{-3}}+A^2\overline{\beta^{-4}}\bigg].
	\label{eq58}
\end{align}
\section{B(E2) Transition rates}
In general, the quadrupole operator is given by \cite{b2}
\begin{eqnarray}
	T_{M}^{(E2)}= t\beta\bigg[ D_{M,0}^{(2)}(\theta_{i})\cos(\gamma)+\frac{1}{\sqrt2} \bigg(D_{M,2}^{(2)}(\theta_{i})  + D_{M,-2}^{(2)}(\theta_{i}) \bigg) \sin(\gamma)\bigg],
	\label{eq59} 
\end{eqnarray}                             
where $t$ is a scaling factor and the Wigner functions of Euler angles are denoted by $D_{M,\alpha}^{(2)}(\theta_{i})$. $(\alpha=0,2,-2)$ represents the angular	momentum quantum number. 
For the $\gamma$-unstable nuclei, $B(E2)$ transition rates from an initial to a final state are given by :
\begin{equation}
	\begin{split}
		B(E2; L_{i}\alpha_{i} \rightarrow L_{f}\alpha_{f})=
		\frac{1}{2L_i+1}	|<\L_{f}\alpha_{f}|T_{M}^{(E2)}|L_{i}\alpha_{i}>|^{2}.
	\end{split}
	\label{eq60}                                                 
\end{equation}
The reduced matrix element $ 	|<\L_{f}\alpha_{f}|T_{M}^{(E2)}|L_{i}\alpha_{i}>|  $ is obtained through the Wigner-Eckart theorem \cite{b50}. Then the B(E2) is given by:
\begin{align}
	B(E2;L_{n,\tau}\rightarrow(L+2)_{n',\tau+1})=\frac{(\tau+1)(4\tau+5)}{(2\tau+5)(4\tau+1)}t^2\times I_{\beta}\bigg(n_{i},L_{i},\alpha_{i},n_{f},L_{f},\alpha_{f}\bigg)^2,
	\label{eq61}
\end{align}
where $I_{\beta}(n_{i},L_{i},\alpha_{i},n_{f},L_{f},\alpha_{f}) $
the integral over $\beta$ having the following form

\begin{eqnarray}
	I_{\beta}(n_{i},L_{i},\alpha_{i},n_{f},L_{f},\alpha_{f})= 
	\int_0^\infty \beta \; \xi_{n_i,L_i,\alpha_i}(\beta) \; \xi_{n_f,L_f,\alpha_f}(\beta)\; \beta^4 \; d\beta,
	\label{eq62}                                                           
\end{eqnarray}

the factor $\beta^4$  comes from the volume element.
\section{ Results and Discussion }
In order to see how the models presented in this paper deal with concrete nuclei, we have used the nuclide chart as well as recent works searching for nuclei with collective spectrum populated in most cases with at least eight  states and whose observed experimental ratios $R_{L/2}$ ($L=0,2,4$) fall in the existence interval of our models. Literally three phenomelogical potentials, namely: infinite square well, Davidson and Kratzer, are used to compare and examine the characteristic structure of these nuclei in the vicinity of the critical point symmetry E(5). The comparison has been done within two formalisms: the minimal length (ML) and the deformation-dependent mass (DDM) for 52 even-even nuclei namely: \ ${}^{98-104}$Ru, ${}^{102-116}$Pd, ${}^{106-120}$Cd, ${}^{118-134}$Xe, ${}^{130-136}$Ba, ${}^{142}$Ba, ${}^{134-138}$Ce,${}^{140}$Nd,${}^{148}$Gd,${}^{140-142}$Sm, ${}^{142-144}$Gd, ${}^{152}$Gd, ${}^{186-200}$Pt. It should be noted that the same set of experimental data for spectra has been previously chosen to be studied in earliear works \cite{b18,b19}, so that these nuclei are good candidate for $\gamma-unstable$.
This section includes all results obtained with E(5)-ML and E(5)-DDM models in four parts \textbf{\ref{sub1}, \ref{sub2}, \ref{sub3}} and \textbf{\ref{sub4}}.
\subsection{Comparison of energy spectra  $R_{L_{g}/n}$ ratios of excited collective states between ML and DDM formalisms for the three potentials}
\label{sub1}
\par The interesting low-lying bands inside the  E(5) symmetry are basically distinguished  by the principal quantum number $n$ and  also the angular momentum $L$. These lowest bands are as follows : the ground state band (g.s.) is characterized by  $n=0$, the $\beta$-band  by  $n=1$ and the $\gamma$-band is obtained from degeneracies of the g.s. levels as follows: the $L=2$ member of the $\gamma$ band  is degenerate  with the $L=4$ member of  g.s. band, the $L=3, 4$  members of the same band are degenerate with the $L=6$ member, the $L=5, 6$ members are degenerate with the $L=8$ member, and so on.
\par The theoretical  energy ratios $R_{4/2}=E(4_1^+)/E(2_1^+)$ of the ground state band, as well as the $\beta$ and $\gamma$ bandheads, normalized to the $2_1^+$ state are denoted by $R_{0/2}=E(0_{\beta}^+)/E(2_1^+)$ and $R_{2/2}=E(2_{\gamma}^+)/E(2_1^+)$ respectively and represented in Tables (\ref{Table1}) and (\ref{Table2}). The numerical results of our elaborated models which are given in Tables (\ref{Table1}) and (\ref{Table2}) are obtained with Eqs.\eqref{eq26}, \eqref{eq30} and \eqref{eq36} for DDM formalism and  Eqs.\eqref{eq42}, \eqref{eq43} and \eqref{eq55} for ML one. It's worth noting that the results presented for DDM have been obtained for $\delta=\lambda=0$. In fact, different choices for $\lambda$ and $\delta$ in the effective potential\eqref{eq10} lead to a renormalization of the optimal parameter values, so that the predicted energy levels remain exactly the same. 
\par An overall impression of the comparison between the theoretical and experimental spectra presented in this work is that the theoretical predictions for our models keep up with the corresponding experimental values for few lower collective states, and after that there is a regression in the agreement between them especially with the Davidson potential. The root mean squared error (r.m.s) is used to evaluate this statement :
\begin{equation}\label{eq63}
	\sigma = \sqrt{ { \sum_{i=1}^n (E_i(exp)-E_i(th))^2 \over
			(n-1)E(2_1^+)^2 } },
\end{equation}
where $E_i(th)$ and $E_i(exp)$  are  the theoretical and experimental energies of the $ i_{th} $ level, respectively, whereas $n$ indicates the number of states. $E(2_1^+) $ is the energy of the first excited level of the g.s band.
\par
\begin{figure*}[htbph]
	\centering	

	\begin{tikzpicture}
		\draw (0, 0) node[inner sep=0] 
		{{\includegraphics[height=65mm]{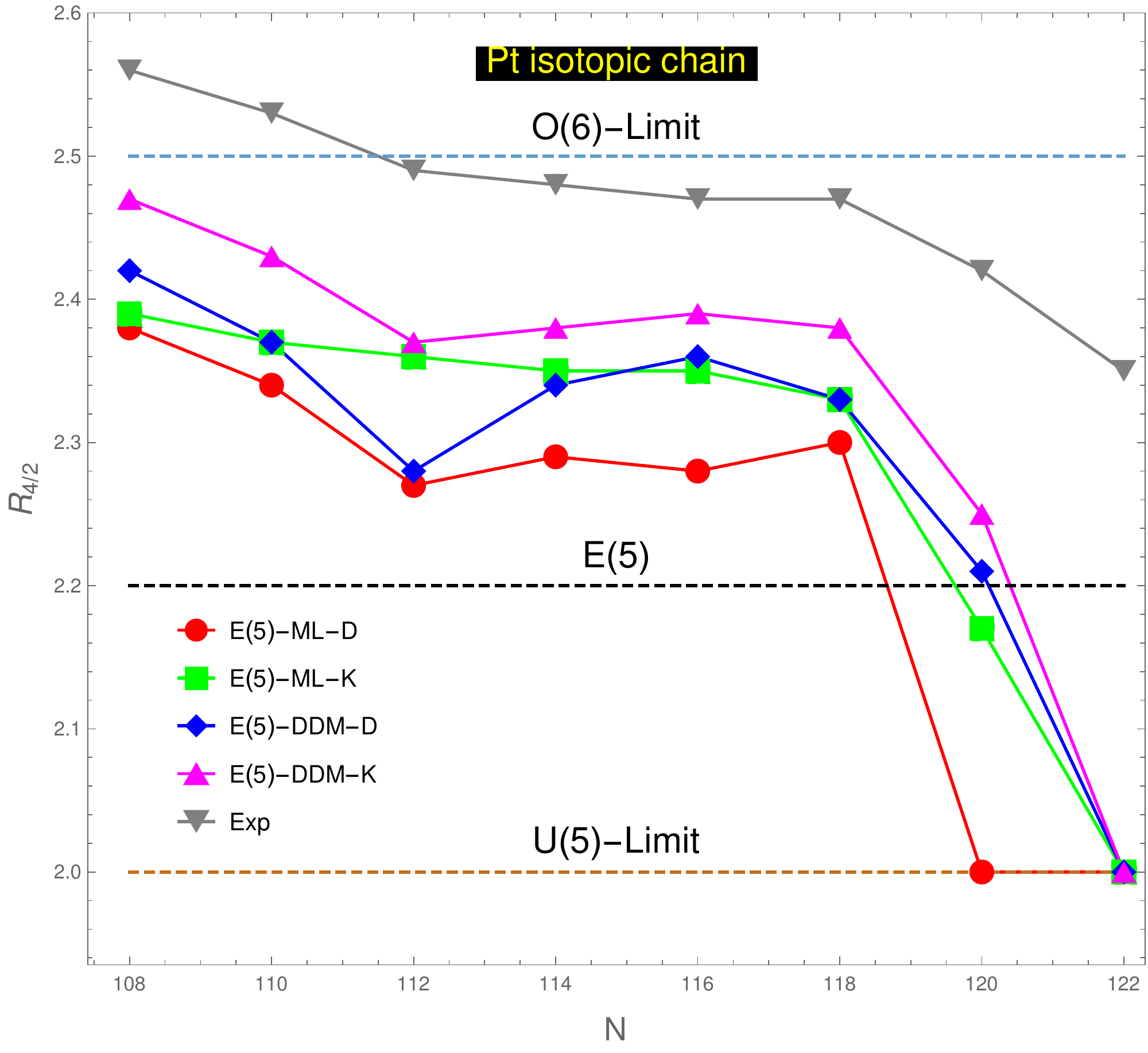}}
			{\includegraphics[height=65mm]{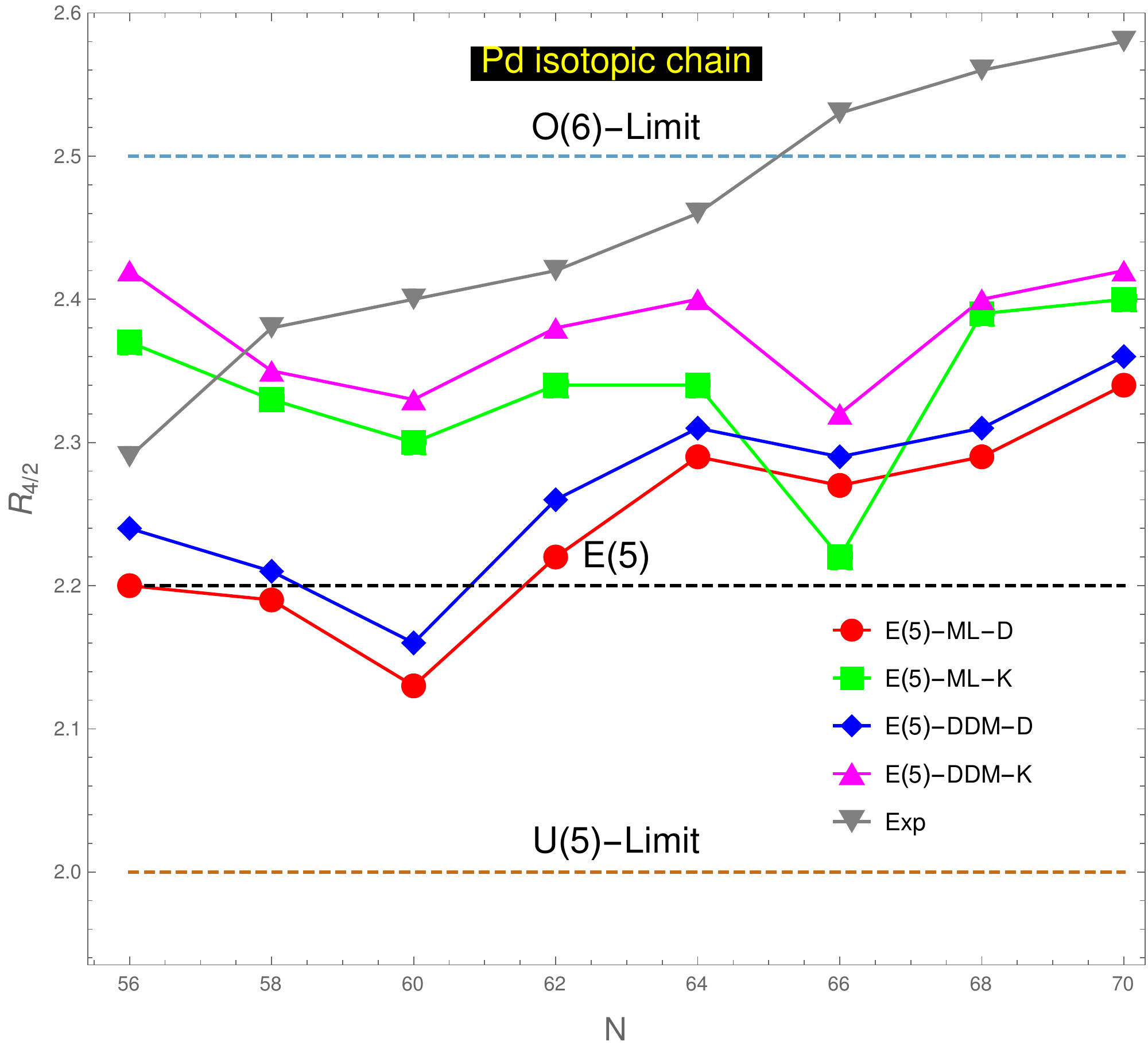}}
		};
		
		\draw (-6, 3) node {\bf (a)};
		\draw (1.3, 3) node {\bf (b)};
		
	\end{tikzpicture}
	\begin{tikzpicture}
		\draw (0, 0) node[inner sep=0] 
		{
			{\includegraphics[height=65mm]{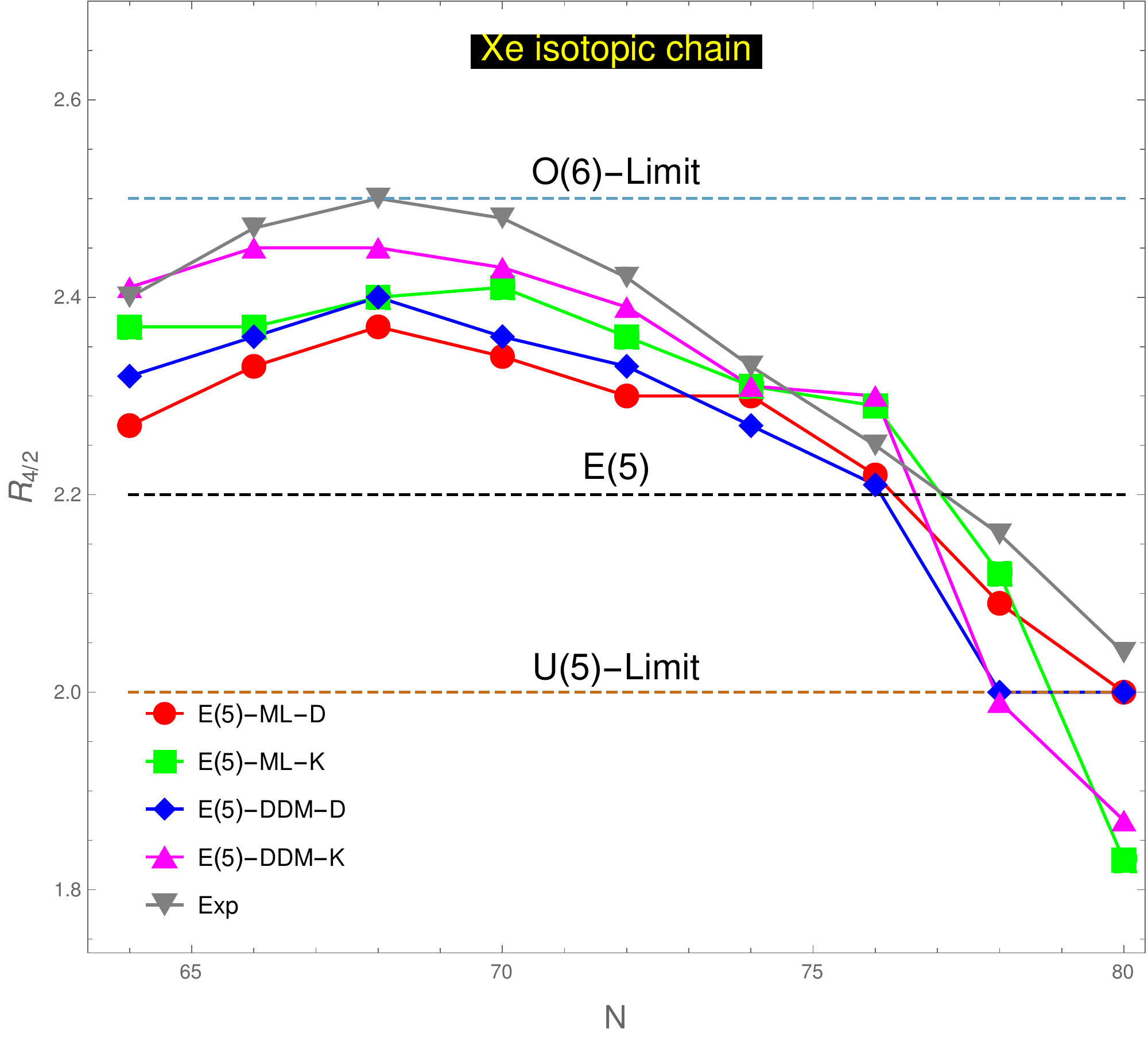}}
			{\includegraphics[height=65mm]{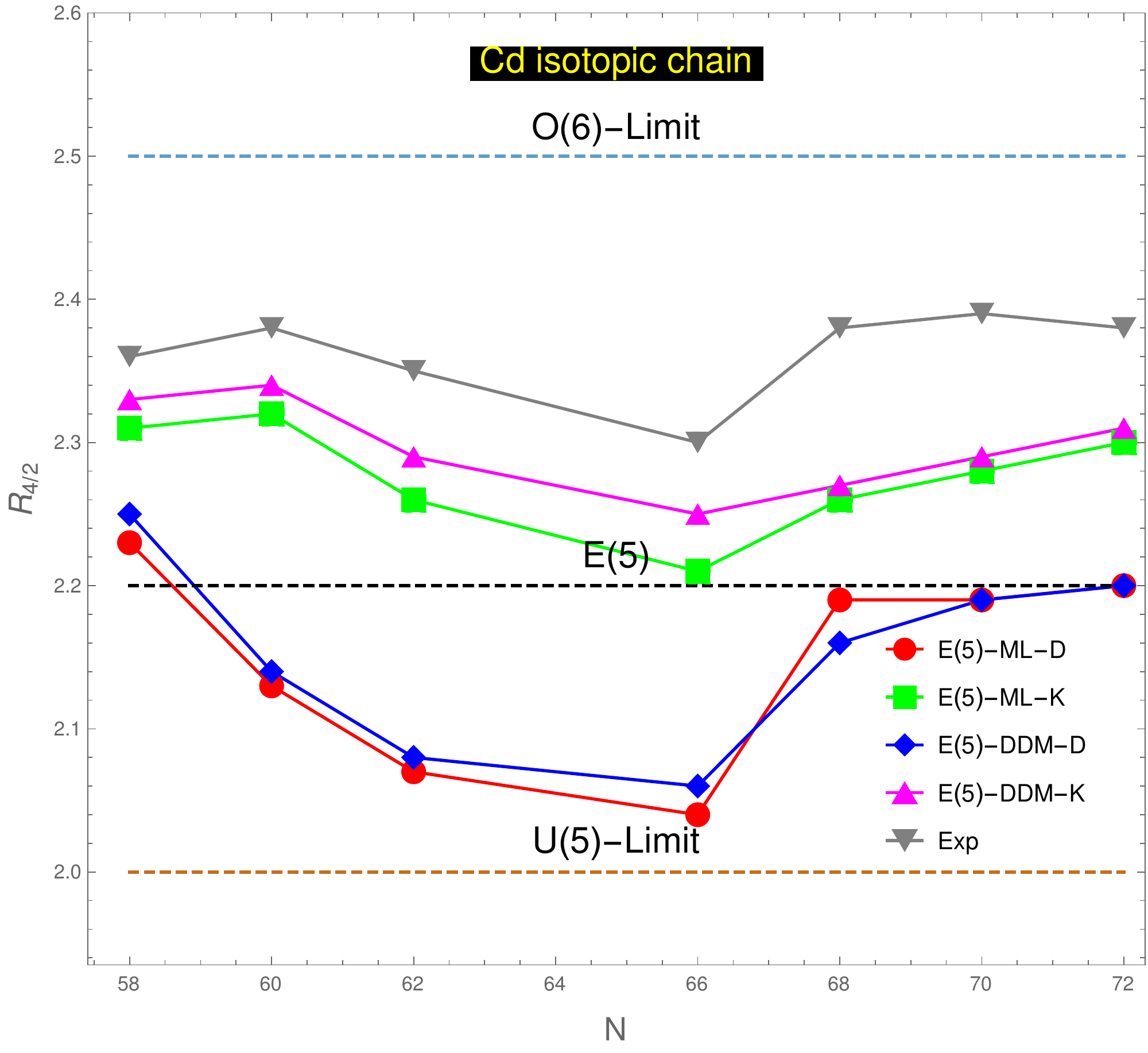}}};
		
		\draw (-6, 3) node {\bf (c)};
		\draw (1.3, 3) node {\bf (d)};
		
	\end{tikzpicture}
	\caption{Evolution of $R_{4/2}$ as function of the neutron number for Pt (upper panel (a)), Pd (upper panel (b)), Xe (lower panel (c)) and Cd (lower panel (d)) isotopic chains. The O(6)-Limit and U(5)-Limit are also shown for comparison.The results presented here are obtained by fitting data for every considered nucleus using the model parameters.}
	\label{Fig1}
\end{figure*}

 Now, we point out the results that we have obtained in this work by involving ML and DDM quantum concepts in E(5), recalling  that E(5) invokes a CPS description of the second-order phase transition along the trajectory from a harmonic vibrator to a $\gamma$-soft rotor. Historically, E(5) was revealed by Iachello even before X(5), the name being derived from the Euclidian algebra in five dimensions. Obviously, the first signature in nuclear structure one would look for is yrast energies consistent with E(5). The $R_{4/2}$ value is $2.19$, intermediate, as would be expected, between the harmonic vibrator ($R_{4/2}= 2.0$) and the $\gamma$-soft rotor ($R_{4/2}= 2.5$). It should be noted, in this context, that the numerical values of the signature ratio $R_{4/2}$, within the framework of each proposed model, depend on the used potentials as we can see in Tables \ref{Table1} and \ref{Table2}.
The evolution of the $R_{4/2}$ ratio as a function of the neutron number for some isotopic chains is depicted in figure (\ref{Fig1}) in the case of the Davidson and Kratzer potentials which thus offers a more flexible description of nuclei. The results are quite interesting.
Furthermore, for all considered nuclei, one can still observe, from the same tables, that the values of both structural parameters $a$ and $\alpha$ do not exceed 0.2, which is consistent with the models' assumption of small deformations as well as the methodologies utilized. Another observation  that can be drawn is that the effect of the DDM and ML formalisms is greater for Davidson's potential than for the others, bearing in mind that for example in our model dubbed E(5)-DDM-D the Davidson  involves two free parameters while in Ref \cite{b18} the used potential contains only one. Our results are therefore much better. The energy adjustment based on two parameters gave results closer to the experimental ones and allowed us to work on more nuclei.
\par Despite the fact that the Davidson and Kratzer potentials have very different shapes especially for high values of $\beta$ , numerical results indicate that a good overall agreement with the experimental data is achieved in both cases especilly for ${}^{106}$Pd, ${}^{106}$Cd, ${}^{108}$Cd, ${}^{128}$Xe and ${}^{134}$Ba. Those nuclei have been already cited in earlier works \cite{b13-a1,b19} to be good candidates for the E(5). 
Depending on the nature of the nuclei (in particular their shapes) as shown in figure (\ref{Fig1}), each potential has a particularity to describe them, especially when the concepts are included. 
However, a global comparison between the results given in both tables and from the values of $\sigma$ one can see that the ML formalism is fairly better for the description of the most nuclei. In addition to the comparison made between the three potentials in each of tables, one can also compare between the two formalisms ML and DDM. Here, one can see that the parameters $a$ and $\alpha$ are  very close to each other for many nuclei.
These remarks lead us to wonder about the relation existing between the parameters of each model and potential. Those parameters are ploted and studied in the following paragraph.
\subsection{Correlation between ML and DDM  formalisms for the three potentials}
\label{sub2}
\begin{figure*}[tbph]
	\centering	
	\begin{tikzpicture}
		\draw (0, 0) node[inner sep=0] 
		{
			{\includegraphics[height=40mm]{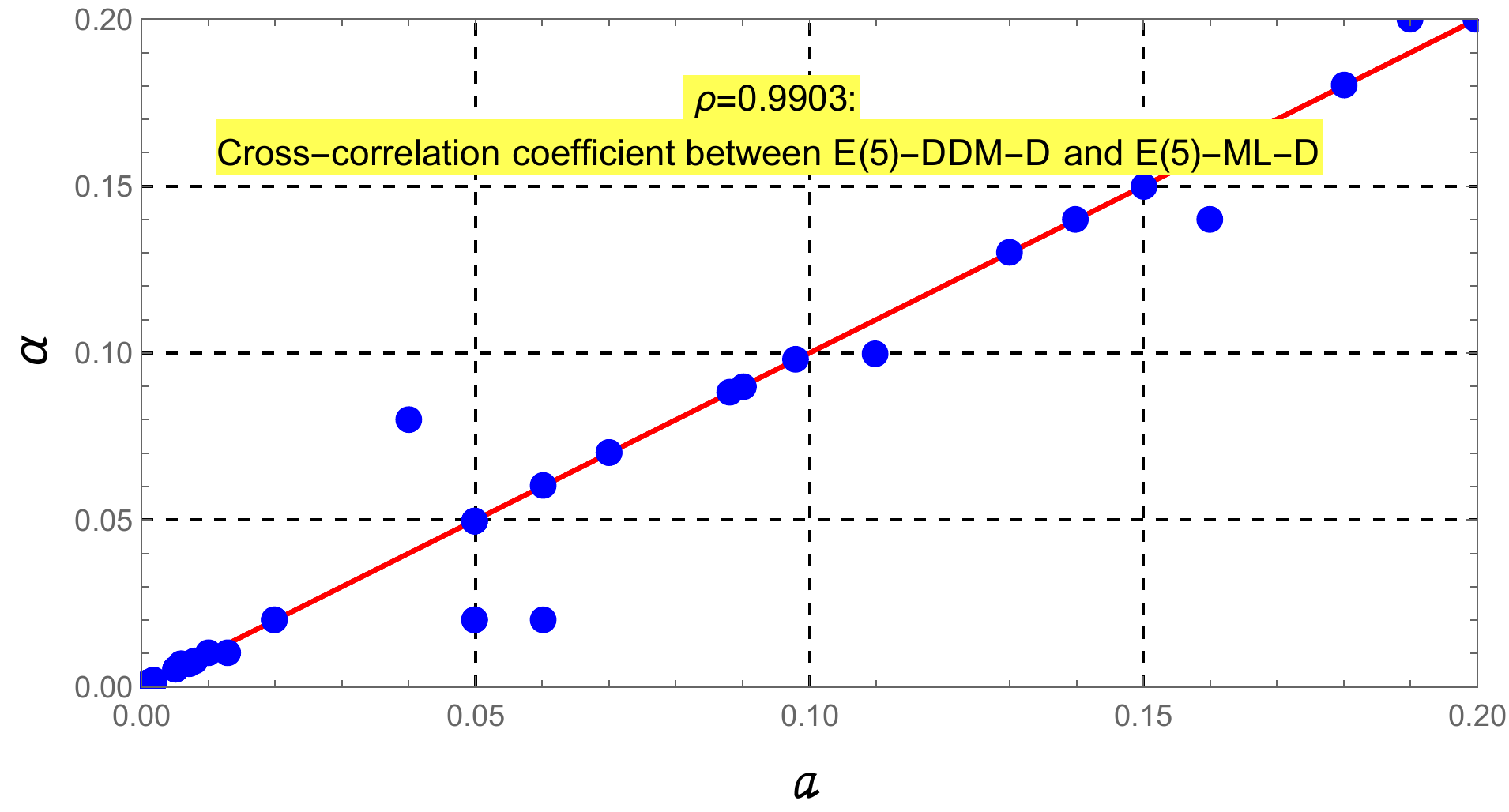}}
			{\includegraphics[height=40mm]{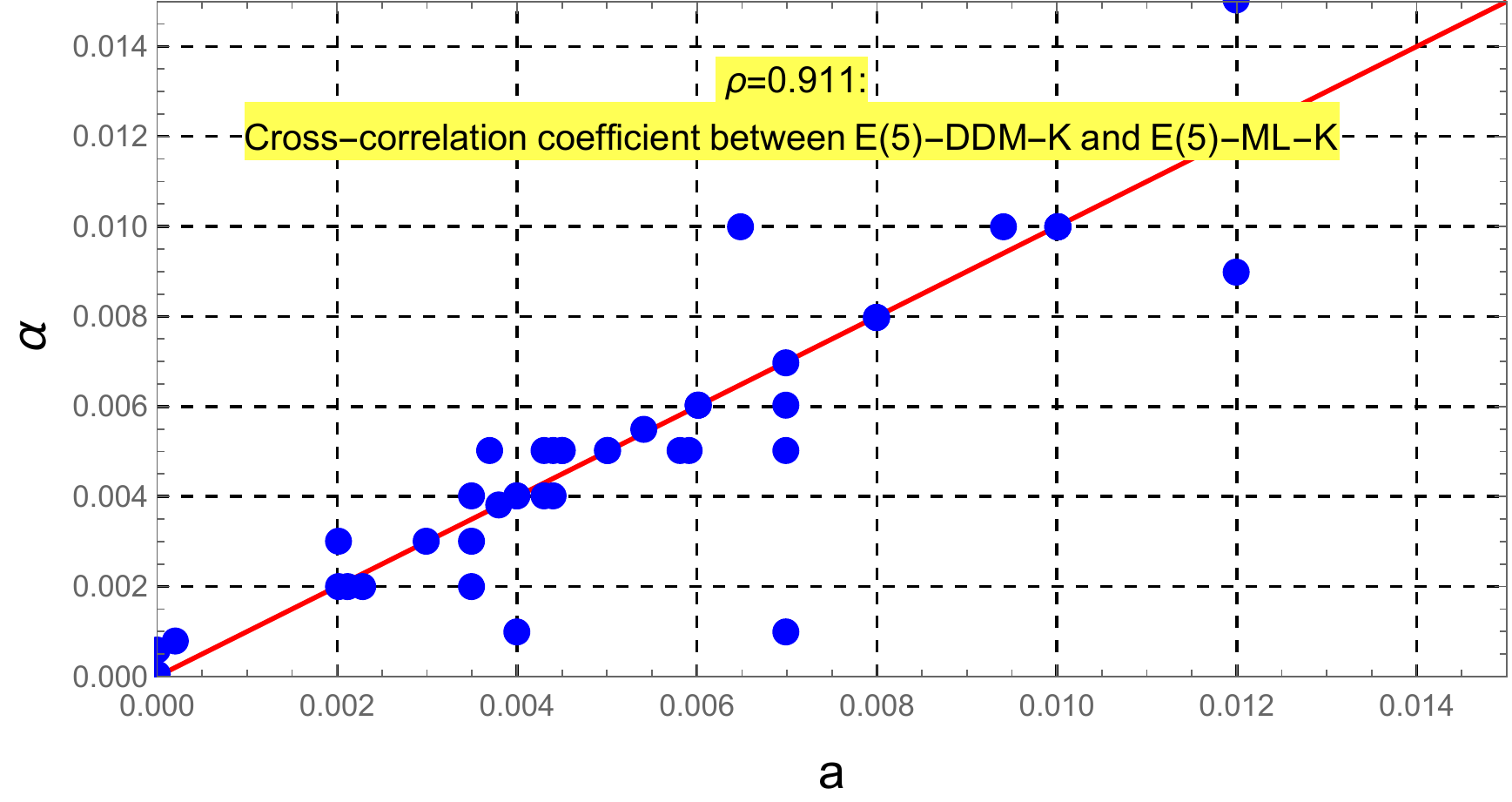}}
		};
		\draw (-6, 1.6) node {\bf (a)};
		\draw (1.3,1.6) node {\bf (b)};	
	\end{tikzpicture}
	\begin{tikzpicture}
		\draw (0, 0) node[inner sep=0] 
		{
			{\includegraphics[height=40mm]{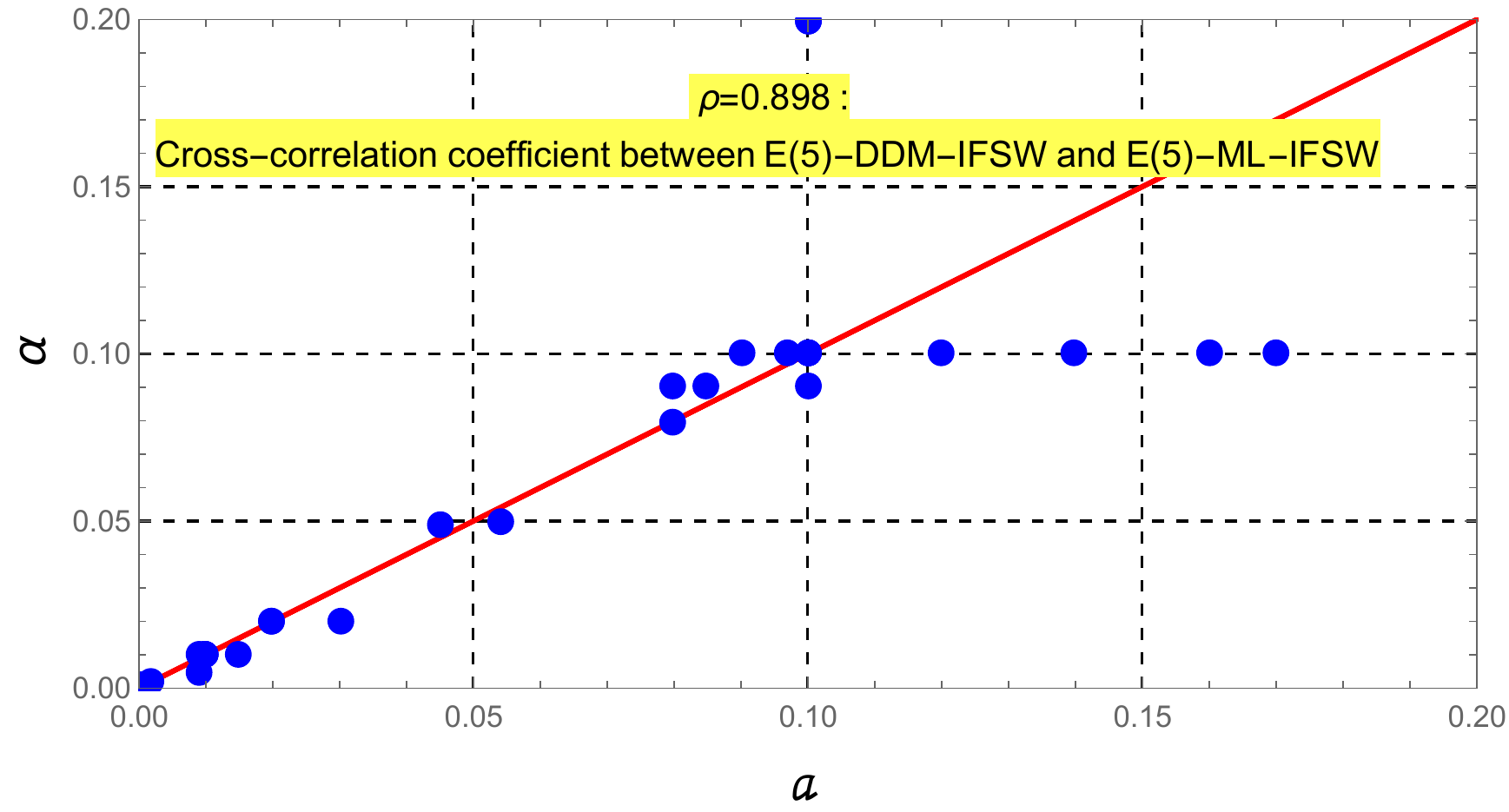}}
		};
		\draw (-2.5, 1.6) node {\bf (c)};
	\end{tikzpicture}		
	\caption{Correlation between the parameters $a$ and $\alpha $ in E(5) model for  Davidson (upper panel(a)), Kratzer (upper panel(b)) and IFSW (lower panel(c)) potentials.}
	\label{Fig2}
\end{figure*}

The parameters' values  $a$ and $\alpha$ that were acquired naturally from the fit on all available experimental data are depicted in Fig.\ref{Fig2}.
Indeed, this figure reveals a strong correlation between ML and DDM concepts where the cross-correlation coefficient for Kratzer and IFSW is close to 90$\%$, whereas for Davidson it is greater than  99$\%$. In the case of IFSW  and Davidson potential, the related  parameters $a$ and $ \alpha$ have values ranging from 0.001 to 0.2. But, in the case of Kratzer potential, they are between 0.001 and 0.17. It is worth noting that several of the derived values of the parameters $a$ and $\alpha$ are extremely near to one another, so the corresponding points on correlation figures are surmounted.  Exactly as in the case of the $\gamma$-rigid regime, the occurrence of such a correlation suggests that the two parameters $a$ and $\alpha$ should not be seen as optional parameters that might take arbitrary values to fit the experimental data, but rather as essential parameters related to the model's structure. 
\subsection{B(E2) transition rates}
\label{sub3}
The quantum observable associated with collectivity is the electromagnetic transition probability which is more directly related to the shape of the nuclear charge distribution. It indicates the presence of coherence in nuclear motion of nuclei. The electromagnetic transition probabilities then provide a very sensitive test of the theoretical models, since they are correlated to the wave functions of the excited states. It's worth noting that  the electric quadrupole excitations (E2) are more prevalent among the other types of excitations. It is for this reason that electric quadrupole transition probability B(E2) represents the most important measurement in the study of the nuclear collectivity. So, in the current study through Tables (\ref{Table3}) and (\ref{Table4}), we provide diverse representative B(E2) transitions normalized to the transition from the first excited level in the ground state band (gsb) and calculated with E(5)-DDM and E(5)-ML for 52 nuclei, employing the same optimal values of the  free parameters derived from fitting the energy spectra for each nucleus. Through the obtained theoretical results, one can easily notice an overall agreement with the experiment for the B(E2) transition rates within the intraband of the gsb for which experimental data are available.
\subsection{Moment of inertia}
\label{sub4}
In this section we must  investigate the influence of the two scenario (DDM and ML) on the variation of the moment of inertia in great detail. There is a little question that such a study will help us to deepen and clarify our current research on the relationship between the two concepts. To begin, it is well understood that the kinematics of nuclei in rotational motion mode is often represented by the moment of inertia, which is traditionally defined as the ratio of the angular momentum L to the angular velocity. However, it may be stated that, for quantum systems, it is very difficult, if not impossible, to offer a universal definition of angular velocity as observable, although there are certain circumstances in which this notion is distinctly defined throughout the process of cranking. The related Inglis model \cite{b52} poses that the particles are surrounded by a distorted rotating self-consistent mean field, which resembles an externally cranked potential. In this regard, the relationship that enables us to determine the moment of inertia of the ground state as a function of the total angular momentum, has the form \cite{b53}:
\begin{equation}
	\theta(L) \approx \frac{2 L-1}{E(L)-E(L-2)},
	\label{eq64}
\end{equation}
which is considerably used to clarify the backbending effect. This is an apparent irregularity in the evolution of angular momentum as a function of the frequency caused by bandcrossing. 
\par
\begin{figure}[tbph]
	\centering
	
	\begin{tikzpicture}
		\draw (0, 0) node[inner sep=0] 
		{	\rotatebox{0}{\includegraphics[height=58mm]{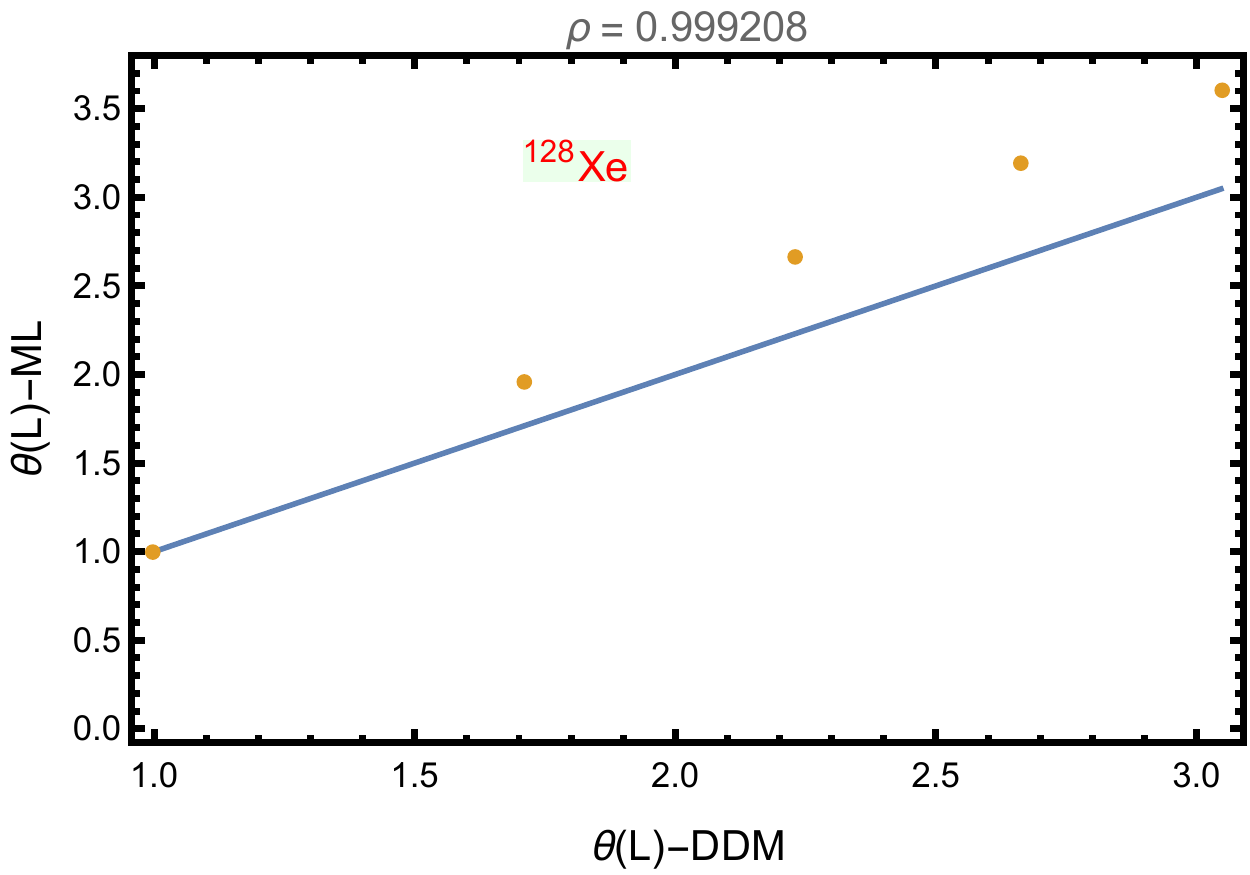}}
			\rotatebox{0}{\includegraphics[height=58mm]{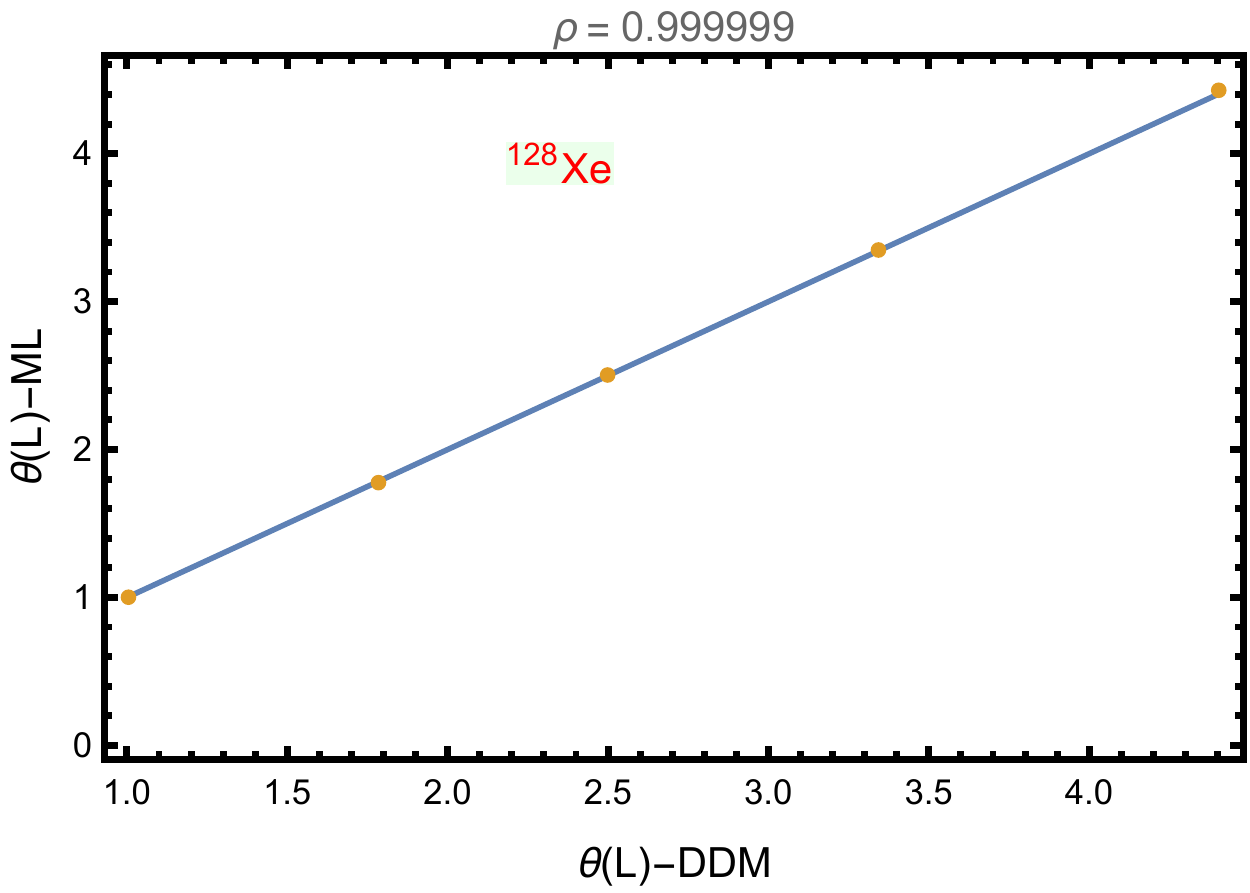}}
		};
		
		\draw (-6.8, 2) node {\bf (a)};
		\draw (1.3, 2) node {\bf (b)};
		
	\end{tikzpicture}
	\begin{tikzpicture}
		\draw (0, 0) node[inner sep=0] 
		{\rotatebox{0}{\includegraphics[height=58mm]{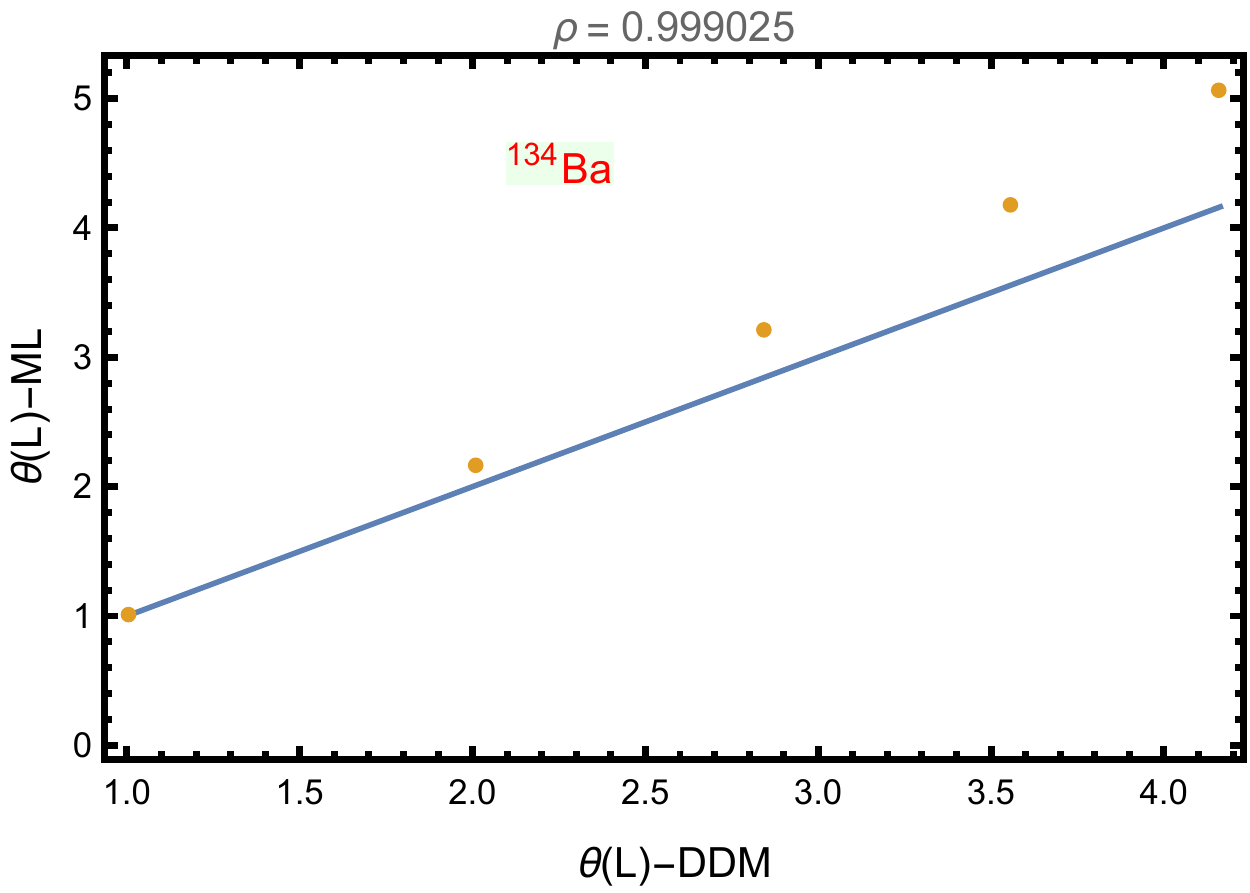}}
			\rotatebox{0}{\includegraphics[height=58mm]{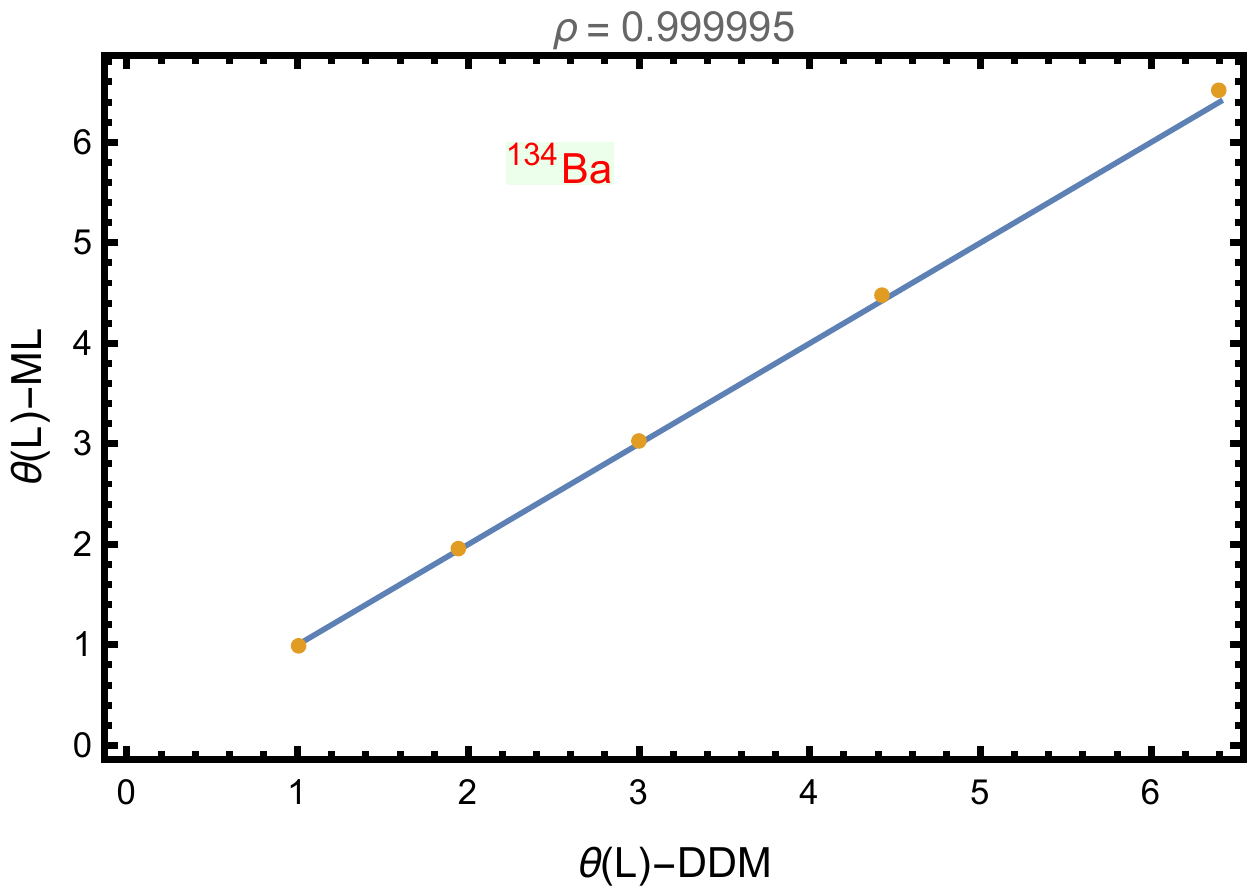}}
		};
		
		\draw (-6.8, 2) node {\bf (c)};
		\draw (1.3, 2) node {\bf (d)};
		
	\end{tikzpicture}
	\caption{The Correlation between moments of inertia for ML and DDM formalism with Davidson (upper panel (a) and (b)) and Kratzer (lower panel (c) and (d)) potentials for $^{128}$Xe  and $^{134}$Ba .}
	\label{Fig3}
\end{figure}

 In order to invistigate the correlation between both concepts, as it has been previously done with other models, we plot the moment of inertia corresponding to both concepts for two candidates ($^{128}$Xe and $^{134}$Ba) of our elaborated models E(5)-ML and E(5)-DDM for the Davidson and Kratzer potentials, as we can see in Fig(\ref{Fig3}).  From this figure, we  can observe  a strong correlation between both concepts, justified by the high values of the cross-correlation coefficient. 
\par To end this part, an important note related to the moment of inertia should be mentioned. So in this regard, we observed that the impacts of both concepts (DDM and ML) on the moment of inertia have  the same behavior for different potential models in the setting of E(5). Indeed, they have a damping effect on the variation of the moment of inertia when the angular momentum increases and keeping it in the norms of the experimental data removing by the way the drawbacks of the model. This finding is already revealed in our pioneering work\cite{b29}, in which we examined the critical point symmetries Z(4) and X(3) under both concepts.
\subsection{Discussion of realization of E(5) CPS in nuclei}
\par The conclusions presented in the final sections of this article are backed up by compelling empirical evidence put forward by a group of authors in their research. This evidence has been gathered and analyzed with meticulous attention to detail using rigorous scientific methods, and it strongly supports our findings that the nuclei located on or close to the bisectrix depicted in figure (\ref{Fig2}) are the most promising candidates. As previously stated, the nuclei that we are specifically referring to are: ${}^{106}\mathrm{Pd}$, ${}^{106}\mathrm{Cd}$, ${}^{108}\mathrm{Cd}$, ${}^{128}\mathrm{Xe}$, and ${}^{134}\mathrm{Ba}$. To exemplify the experimental evidence, Casten and Zamfir\cite{Casten} proposed ${ }^{134} \mathrm{Ba}$ as a possible experimental candidate for an $\mathrm{E}(5)$ nucleus in their research. Although there were certain limitations in comparing the absolute transition probabilities with theoretical calculations, they highlighted that ${}^{134}\mathrm{Ba}$ is close to exemplifying E(5) symmetry\cite{Casten}.
In the same context, Zamfir et al.\cite{Zamfir} proposed ${ }^{102} \mathrm{Pd}$ as another possible candidate for an $\mathrm{E}(5)$  nucleus, but the lifetimes reported by Konstantinopoulos et al.\cite{Konstan} do not seem to support this suggestion. To conduct a more systematic search for possible candidates exhibiting $\mathrm{E}(5)$  characteristics, Clark et al.\cite{Clark} scoured the ENSDF database\cite{b54} for nuclei that met certain criteria.
The first step was to look for transitional nuclei in the mass regions of $30 \leqslant Z \leqslant 82$ and $A \geqslant 60$, with $2.00<$ $E\left(4_{1}^{+}\right) / E\left(2_{1}^{+}\right) \leqslant 2.40$, which yielded over 70 possible candidates. The second requirement was for the existence of two excited $0^{+}$ states within 2.5 and 4.5 times $E\left(2_{1}^{+}\right)$. Only six nuclei passed both criteria: ${ }^{102} \mathrm{Pd},{ }^{106,108} \mathrm{Cd}$, ${ }^{124} \mathrm{Te},{ }^{128} \mathrm{Xe}$, and ${ }^{134} \mathrm{Ba}$. After examining the available data against the remaining criteria regarding the decays of the excited $0^{+}$ states, only ${ }^{128} \mathrm{Xe}$ and ${ }^{134} \mathrm{Ba}$ were identified as viable candidates. Since ${ }^{134} \mathrm{Ba}$ had already been suggested as a candidate by Casten and Zamfir\cite{Casten}, ${ }^{128} \mathrm{Xe}$ was the sole newly identified $\mathrm{E}(5)$ candidate.
Another study conducted by Coquard et al.\cite{Coquard} in 2009 used Coulomb excitation in inverse kinematics to analyze ${ }^{128} \mathrm{Xe}$ and obtain detailed spectroscopic information, including $B(E2)$ values for numerous transitions. Based on their analysis of the relative energies of the excited $0^{+}$ states and the absolute $B(E2)$ values for their decays, the researchers concluded that ${ }^{128} \mathrm{Xe}$ is not an $\mathrm{E}(5)$ nucleus and proposed that ${ }^{130} \mathrm{Xe}$ might be a more suitable candidate. However, in subsequent Coulomb excitation measurements on ${ }^{130,132} \mathrm{Xe}$ by the same group, no evaluation was made of the $\mathrm{E}(5)$ character of these isotopes, likely because the excited $0^{+}$ states were not significantly populated in either nucleus. Peters et al.\cite{Peters} also recently suggested that the same nuclei, i.e., ${}^{130,132}\mathrm{Xe}$, have similar nuclear structures that exhibit the E(5) critical point symmetry. They investigated the level structures of these nuclei using the inelastic neutron scattering reaction, followed by $\gamma$-ray detection. The researchers measured the level lifetimes of the nuclei with the Doppler-shift attenuation method and characterized their low-lying excited states. It is worth mentioning that Liu and Zhang\cite{Liu} also proposed the use of ${ }^{130} \mathrm{Xe}$. Additionally, Frank et al.\cite{Frank} and Long et al.\cite{Long} independently suggested the ${ }^{104} \mathrm{Ru}$ and ${ }^{114} \mathrm{Cd}$ nuclei, respectively, which are listed in Tables \ref{Table1} and \ref{Table2}. 
\par In certain instances, when comparing the results of a theoretical model to experimental data, it may not be possible to achieve a specific level of precision. Nonetheless, this outcome should not be interpreted as a failure, but rather as an indication that other factors may be involved.
\section{Conclusion}
Through this study, we have tested  the robustness of the E(5) solution under ML and DDM concepts, for some concrete nuclei, by addressing the role of IFSW, Davidson and Kratzer potentials on some quantities as energy spectra,  electric quadrupole transitions and moment of inertia. After successful reproduction of the experimental energy spectra and the electromagnetic B(E2) transition probabilities, the correlation between the two quantum concepts (ML and DDM) for some nuclei in the vicinity of the E(5) critical point symmetry such as X(3) and Z(4) \cite{b29} has been observed. The present study comes to corroborate the fact that the revealed correlation between both quantum concepts (ML and DDM) is universal and does not depend on the used model.
\begin{table*}[tbph]
	\centering
	{\renewcommand{\arraystretch}{1}
		{\setlength{\tabcolsep}{0.5cm} 
			\caption{Comparison of experimental data \cite{b54} (upper lines) of the $\gamma$-unstable Bohr Hamiltonian to theoretical predictions with ML formalism (lower lines) for Davidson (D), Kratzer (K) and Infinite Square Well (ISW) potential respectevely.}
			\label{Table1}
			\resizebox{0.90\textwidth}{!}{%
				\begin{tabular}{|ccccccccccc|}		
					\hline 
					Nucleus &&$ {R_{4,2}}$ & $ {R_{0,2}}$&$ {R_{2,2}}$&$ \alpha $&c&$ \beta_{0} $&$\tilde{B}$&$\sigma_{r.m.s.}$&\\
					\hline
					{}	&&&&&&&&&&\\
					{$^{98}$Ru} &{\scriptsize Exp}&2.14&2.0&2.2&&&&&&\\
					&	{\scriptsize D}&2.13&2.45&2.1&0.0100&5.00&1.42&-&0.26&\\
					&	{\scriptsize K}&2.33&3.95&2.3&0.0100&-&-&74&0.90 &\\
					&{\scriptsize IFSW}&1.49&1.68&1.4&0.0121&-&-&-&0.87&\\
					
					{$^{100}$Ru} &{\scriptsize Exp}&2.27&2.1&2.5 &&&&&&\\
					&	{\scriptsize D}&2.32&2.32&2.3&0.0077&199&0&-&0.41& \\
					&	{\scriptsize K}&2.33&3.94&2.3&0.0090&-&-&74&0.82& \\
					&{\scriptsize IFSW}&2.21&3.07&2.2&0.0199&-&-&-&0.70& \\
					{$^{102}$Ru} &{\scriptsize Exp}&2.33&2.0&2.3 &&&&&&\\
					&	{\scriptsize D}&2.18&2.43&2.1&0.0100&3.37&0.12&-&0.35 &\\
					&	{\scriptsize K}&2.33&4.03&2.3&0.0100&-&-&76&0.68&\\
					&{\scriptsize IFSW}&2.20&3.04&2.2&0.0124&-&-&-&0.44&\\
					{$^{104}$Ru} &{\scriptsize Exp}&2.48&2.8&2.5 &&&&&&\\
					&	{\scriptsize D}&2.34&2.75&2.3&0.0044&39.89&0.21&-&0.34& \\
					&	{\scriptsize K}&2.31&3.76&2.3&0.0010&-&-&67&0.57&\\
					&{\scriptsize IFSW}&2.19&3.03&2.2&0.01999&-&-&-&0.89 &\\
					{$^{102}$Pd}&{\scriptsize Exp}&2.29&2.9&2.8& &&&&&\\
					&{\scriptsize D}& 2.20&2.34&2.1&0.0020&19.91&0.073&-&0.45& \\
					&	{\scriptsize K}&2.37&4.71&2.3&0.0010&-&-&105&0.86&\\
					&{\scriptsize IFSW}&2.19 &3.03&2.2&0.0010&-&-&-&0.98& \\
					{$^{104}$Pd} &{\scriptsize Exp}&2.38 &2.4&2.4 &&&&&&\\
					&{\scriptsize D}&2.19&2.88 &2.19&0.0076&12.44&0.36&-&0.44& \\
					&	{\scriptsize K}&2.33&4.01&2.3&0.0050&-&-&77&0.64& \\
					&{\scriptsize IFSW}&2.19&2.03 &2.20&0.0050&-&-&-&0.86&  \\
					{$^{106}$Pd} &{\scriptsize Exp}&2.40&2.2 &2.2&&&&&& \\
					&{\scriptsize D}&2.13&2.15&2.1&0.0050&6.34&0&-&0.45& \\
					&	{\scriptsize K}&2.30&3.63&2.3&0.0080&-&-&63&0.75& \\
					&{\scriptsize IFSW}&1.49&1.68&1.4&0.0010&-&-&-&0.85&  \\
					{$^{108}$Pd}&{\scriptsize Exp}&2.42&2.4 &2.1 &&&&&&\\
					&{\scriptsize D}&2.22&2.23 &2.2&0.0050&16.06&0&-&0.38& \\
					&	{\scriptsize K}&2.34&4.05&2.3&0.0060&-&-&78&0.90 &\\
					&{\scriptsize IFSW}&2.19& 3.03&2.19&0.0010&-&-&-&0.87& \\
					{$^{110}$Pd}&{\scriptsize Exp}&2.46&2.5 &2.2 &&&&&&\\
					&{\scriptsize D}&2.29&2.30 &2.3&0.0076&39.99&0&-&0.49& \\
					&	{\scriptsize K}&2.34&4.13&2.3&0.0060&-&-&81&1.00& \\
					&{\scriptsize IFSW}&2.19& 3.03&2.2&0.0012&-&-&-&0.93& \\
					{$^{112}$Pd}&{\scriptsize Exp}&2.53 &2.6&2.1 &&&&&&\\
					&{\scriptsize D}& 2.27&2.53&2.2&0.0019&38.86&0.13&-&0.39 &\\
					&	{\scriptsize K}&2.22&2.86&2.2&0.0050&-&-&38&0.46 &\\
					&{\scriptsize IFSW}&2.20&3.04 &2.2&0.0099&-&-&-&0.46&  \\
					{$^{114}$Pd} &{\scriptsize Exp}&2.56 &2.6&2.1& &&&&&\\
					&{\scriptsize D}&2.29&2.75 &2.2&0.0100&6.98&0.26&-&0.60& \\
					&	{\scriptsize K}&2.39&5.04&2.3&0.0100&-&-&120&0.90 &\\
					&{\scriptsize IFSW}&2.22&3.1 &2.2&0.0745&-&-&-&0.64&\\
					\hline 
	\end{tabular}}}}
\end{table*}
\begin{table*}[tbph]
	\centering
	\setcounter{table}{0} \caption{(continued)}
	{\renewcommand{\arraystretch}{1}
		{\setlength{\tabcolsep}{0.5cm} 
			\label{Table1p}
			\resizebox{0.90\textwidth}{!}{%
				\begin{tabular}{|ccccccccccc|}		
					\hline 
					Nucleus &&$ {R_{4,2}}$ & $ {R_{0,2}}$&$ {R_{2,2}}$&$ \alpha $&c&$ \beta_{0} $&$\tilde{B}$&$\sigma_{r.m.s.}$&\\
					\hline
					{}	&&&&&&&&&&\\
					{$^{116}$Pd}&{\scriptsize Exp}&2.58&3.3 &2.2 &&&&&&\\
					&{\scriptsize D}&2.34&3.45 &2.3&0.0024&14.38&0.57&-&0.57& \\
					&	{\scriptsize K}&2.4&5.59&2.4&0.0050&-&-&146&0.93& \\
					&{\scriptsize IFSW}&2.24& 3.14&2.2&0.0843&-&-&-&0.66 &\\
					{$^{106}$Cd}&{\scriptsize Exp}&2.36&2.8&2.7&& &&&&\\
					&{\scriptsize D}&2.23&2.97&2.2&0.0030&0.003&0.51&-&0.24& \\
					&	{\scriptsize K}&2.31&3.65&2.3&0.0040&-&-&64&0.35& \\
					&{\scriptsize IFSW}&2.19&3.03&2.1&0.0010&-&-&-&0.32 & \\
					{$^{108}$Cd}  &{\scriptsize Exp}&2.38&2.7&2.52&&& &&&\\
					&{\scriptsize D}&2.13&2.37&2.1&0.0099&1.96&0.12&-&0.48 &\\
					&	{\scriptsize K}&2.32&3.8&2.3&0.0050&-&-&69&0.69& \\
					&{\scriptsize IFSW}&2.19&3.03&2.2 &0.0010&-&-&-&0.46& \\
					{$^{110}$Cd} &{\scriptsize Exp}&2.35&2.2&2.2& &&&&&\\
					&{\scriptsize D}&2.07&2.13&2.0 &0.0100&2.13&0&-&0.54&\\
					&	{\scriptsize K}&2.26&3.18&2.2&0.0100&-&-&48&0.74& \\
					&{\scriptsize IFSW}&2.19&3.03&2.1&0.0010&-&-&-&0.76&  \\
					{$^{114}$Cd} &{\scriptsize Exp}&2.30&2.0&2.2& &&&&&\\
					&{\scriptsize D}&2.04&2.07&2.0 &0.0050&2.29&0&-&0.46&\\
					&	{\scriptsize K}&2.21&2.8&2.2 &0.0090&-&-&37&0.59&\\
					&{\scriptsize IFSW}&2.19&3.03&2.1&0.0010&-&-&-& 0.92& \\
					{$^{116}$Cd}  &{\scriptsize Exp}&2.38&2.5&2.4&&&&&&\\
					&{\scriptsize D}&2.19&2.68&2.1&0.0010&9.07&0.33&-&0.32 & \\
					&	{\scriptsize K}&2.26&3.18&2.2&0.0030&-&-&48&0.32& \\
					&{\scriptsize IFSW}&2.19&3.03&2.&0.0020&-&-&-&0.79& \\
					\hline 
	\end{tabular}}}}
\end{table*}
\begin{table*}[tbph]
	\centering
	\setcounter{table}{0} \caption{(continued)}
	{\renewcommand{\arraystretch}{1}
		{\setlength{\tabcolsep}{0.5cm} 
			\label{Table3a}
			\resizebox{0.90\textwidth}{!}{%
				\begin{tabular}{|cccccccccc|}		
					\hline 
					Nucleus &&$ {R_{4,2}}$ & $ {R_{0,2}}$&$ {R_{2,2}}$&$ \alpha $&c&$ \beta_{0} $&$ \tilde{B} $&$\sigma_{r.m.s.}$\\
					\hline
					{}	&&&&&&&&&\\
					{$^{118}$Cd} &{\scriptsize Exp}&2.39&2.6& 2.6&&&&&\\
					&{\scriptsize D}&2.19&2.94&2.1&0.0020&0.39&0.55&-&0.38\\
					&	{\scriptsize K}&2.28&3.36&2.2&0.0050&-&-&54&0.36 \\
					&{\scriptsize IFSW}&2.19&3.03&2.2&0.0150&-&-&-&0.90  \\
					{$^{120}$Cd} &{\scriptsize Exp}&2.38&2.7& 2.6&&&&&\\
					&{\scriptsize D}&2.20&2.95&2.2&0.0020&1.27&0.55&-&0.37\\
					&	{\scriptsize K}&2.30&3.61&2.3&0.0050&-&-&62&0.47 \\
					&{\scriptsize IFSW}&2.19&3.03&2.2&0.0100&-&-&-&0.90 \\
					{$^{118}$Xe} &{\scriptsize Exp}&2.40&2.5& 2.8&&&&&\\
					&{\scriptsize D}&2.27&2.43&2.2&0.0050&26.70&0.04&-&0.39\\
					&	{\scriptsize K}&2.37&4.68&2.3&0.0050&-&-&104&0.92 \\
					&{\scriptsize IFSW}&2.21&3.02&2.2&0.0199&-&-&-&0.94 \\
					{$^{120}$Xe} &{\scriptsize Exp}&2.47&2.8& 2.7&&&&&\\
					&{\scriptsize D}&2.33&3.39&2.31&0.0010&30.33&0.70&-&0.58\\
					&	{\scriptsize K}&2.37&4.71&2.3&0.0010&-&-&69&0.60 \\
					&{\scriptsize IFSW}&2.24&3.14&2.2&0.0840&-&-&-&0.69 \\
					{$^{122}$Xe} &{\scriptsize Exp}&2.50&3.5& 2.5&&&&&\\
					&{\scriptsize D}&2.37&3.38&2.37&0.1999&1.92&0.53&-&0.58\\
					&	{\scriptsize K}&2.40&5.35&2.4&0.1900&-&-&134&0.81 \\
					&{\scriptsize IFSW}&2.30&3.3&2.30&0.1749&-&-&-& 0.33 \\	{$^{124}$Xe} &{\scriptsize Exp}&2.48&3.6& 2.4&&&&&\\
					&{\scriptsize D}&2.34&3.60&2.3&0.0040&6.85&0.91&-&0.54\\
					&	{\scriptsize K}&2.41&5.68&2.4&0.0040&-&-&134&0.91 \\
					&{\scriptsize IFSW}&2.22&3.08&2.2&0.0717&-&-&-& 0.72 \\	
					{$^{126}$Xe}&{\scriptsize Exp}&2.42&3.4& 2.3&&&&&\\
					&{\scriptsize D}&2.30&2.95&2.3&0.0030&15.41&0.39&-&0.57\\
					&	{\scriptsize K}&2.36&4.41&2.3&0.0030&-&-&92&0.78 \\
					&{\scriptsize IFSW}&2.21&3.06&2.2&0.0149&-&-&-&0.81\\	
					{$^{128}$Xe} &{\scriptsize Exp}&2.33&3.6& 2.2&&&&&\\
					&{\scriptsize D}&2.30&3.44&2.3&0.0000&398&0.82&-&0.22\\
					&	{\scriptsize K}&2.31&2.7&2.3&0.0022&-&-&63&0.41 \\
					&{\scriptsize IFSW}&2.19&3.03&2.20&0.0012&-&-&-&0.42 \\
					{$^{130}$Xe} &{\scriptsize Exp}&2.25&3.3& 2.1&&&&&\\
					&{\scriptsize D}&2.22&3.14&2.2&0.0000&1.79&0.71&-&0.18\\
					&	{\scriptsize K}&2.29&3.49&2.29&0.0060&-&-&58&0.42 \\
					&{\scriptsize IFSW}&2.19&3.03&2.2&0.0010&-&-&-& 0.67 \\
					{$^{132}$Xe}&{\scriptsize Exp}&2.16&2.8&1.9&&&&& \\
					&{\scriptsize D}&2.09&2.42&2.1&0.0029&0.22&0.21&-&0.25\\
					&	{\scriptsize K}&2.12&2.37&2.1&0.0049&-&-&25&0.26 \\
					&{\scriptsize IFSW}&2.19&3.03&2.2&0.0010&-&-&-& 0.83\\
					{$^{134}$Xe}  &{\scriptsize Exp}&2.04&1.9&1.9&&&&& \\
					&{\scriptsize D}&2.00&2.00&2.0&0.0000&0.25&0&-&0.59 \\
					&	{\scriptsize K}&1.83&1.57&1.8&0.0000&-&-&9&0.17 \\
					&{\scriptsize IFSW}&1.48&1.06&1.4&0.0000&-&-&-&0.54  \\
					\hline 
	\end{tabular}}}}
\end{table*}  
\begin{table*}[tbph]
	\centering
	\setcounter{table}{0} \caption{(continued)}
	{\renewcommand{\arraystretch}{1}
		{\setlength{\tabcolsep}{0.5cm} 
			\label{Table3a}
			\resizebox{0.90\textwidth}{!}{%
				\begin{tabular}{|cccccccccc|}		
					\hline 
					Nucleus &&$ {R_{4,2}}$ & $ {R_{0,2}}$&$ {R_{2,2}}$&$ \alpha $&c&$ \beta_{0} $&$ \tilde{B} $&$\sigma_{r.m.s.}$\\
					\hline
					{}	&&&&&&&&&\\
					{$^{130}$Ba} &{\scriptsize Exp}&2.52&3.3&2.5&&&&& \\
					&{\scriptsize D}&2.40&3.31&2.4 &0.0099&39.99&0.52&-&0.31\\
					&	{\scriptsize K}&2.4&5.4&2.4&0.0050&-&-&137&0.88 \\
					&{\scriptsize IFSW}&2.21&3.06&2.2&0.000&-&-&-&0.37  \\
					{$^{132}$Ba} &{\scriptsize Exp}&2.43&3.2&2.2&&&&& \\
					&{\scriptsize D}&2.40&3.31&2.4 &0.0099&4.14&0.48&-&0.38\\
					&	{\scriptsize K}&2.36&4.51&2.3 &0.005&-&-&96&0.70\\
					&{\scriptsize IFSW}&2.22&3.09&2.2&0.0099&-&-&-& 0.37 \\
					{$^{134}$Ba}  &{\scriptsize Exp}&2.32&2.9&1.9&&&&&\\
					&{\scriptsize D}&2.15&2.69&2.1&0.0019&0.20&0.38&-&0.30 \\
					&	{\scriptsize K}&2.19&2.73&2.1&0.0006&-&-&35&0.29 \\
					&{\scriptsize IFSW}&2.19&3.03&2.2&0.001&-&-&-& 0.38\\
					{$^{136}$Ba} &{\scriptsize Exp}&2.28&1.9& 1.9&&&&&\\
					&{\scriptsize D}&2.00&2.07&2.0&0.0010&0.25&0&-&0.19\\
					&	{\scriptsize K}&1.98&1.89&1.9&0.0008&-&-&15&0.14 \\
					&{\scriptsize IFSW}&2.19&3.03&2.2&0.0010&-&-&-&0.65 \\
					{$^{142}$Ba} &{\scriptsize Exp}&2.32&4.27& 3.96&&&&&\\
					&{\scriptsize D}&2.37&4.34&2.37&0.0023&5.72&1.66&-&0.54\\
					&	{\scriptsize K}&2.40&5.38&2.4&0.0020&-&-&136&0.67 \\
					&{\scriptsize IFSW}&2.24&3.13&2.2&0.0199&-&-&-&0.76 \\
					{$^{134}$Ce} &{\scriptsize Exp}&2.56&3.7& 2.4&&&&&\\
					&{\scriptsize D}&2.34&4.21&2.34&0.0010&4.72&1.62&-&0.57\\
					&	{\scriptsize K}&2.40&5.57&2.4&0.0030&-&-&145&0.83 \\
					&{\scriptsize IFSW}&2.21&3.06&2.2&0.0199&-&-&-&0.85 \\	
					\hline 
	\end{tabular}}}}
\end{table*}  

\begin{table*}[tbph]
	\centering
	\setcounter{table}{0} \caption{(continued)}
	{\renewcommand{\arraystretch}{0.95}
		{\setlength{\tabcolsep}{0.5cm} 
			\label{Table1c}
			\resizebox{0.90\textwidth}{!}{%
				\begin{tabular}{|ccccccccccc|}		
					\hline 
					Nucleus &&$ {R_{4,2}}$ & $ {R_{0,2}}$&$ {R_{2,2}}$&$ \alpha $&c&$ \beta_{0} $&$ \tilde{B} $&$\sigma_{r.m.s.}$&\\
					\hline
					{}	&&&&&&&&&&\\
					{$^{142}$Sm}  &{\scriptsize Exp}&2.33&1.9&2.2&&&&&& \\
					&{\scriptsize D}&2.05&2.06&2.08&0.0050&2.63&0&-&0.14& \\
					&	{\scriptsize K}&2.11&2.31&2.1&0.0080&-&-&24&0.22& \\
					&{\scriptsize IFSW}&2.19&3.03&2.1&0.0010&-&-&-&0.60& \\
					{$^{142}$Gd} &{\scriptsize Exp}&2.35&2.7&1.9& &&&&&\\
					&{\scriptsize D}&2.20&2.84&2.2 &0.005&2.03&0.43&-&0.23&\\
					&	{\scriptsize K}&2.32&3.78&2.3&0.0050&-&-&68&0.42 &\\
					&{\scriptsize IFSW}&2.19&3.03&2.1&0.001&-&-&-&0.87& \\
					{$^{144}$Gd} &{\scriptsize Exp}&2.35&2.5&2.5&&&&&& \\
					&{\scriptsize D}&2.31&2.54&2.3 &0.0094&19.98&0.11&-&0.10&\\
					&	{\scriptsize K}&2.21&2.81&2.2 &0.0060&-&-&37&0.26&\\
					&{\scriptsize IFSW}&2.20&3.04&2.2&0.0099&-&-&- &0.29& \\
					{$^{152}$Gd}  &{\scriptsize Exp}&2.19&1.8&3.2&&&&&&\\
					&{\scriptsize D}&2.22&2.24&2.2&0.0100&8.91&0&-&0.49&  \\
					&	{\scriptsize K}&2.28&3.33&2.3&0.0090&-&-&53&0.87&\\
					&{\scriptsize IFSW}&2.19&3.03&2.2&0.0011&-&-&-&0.64&  \\
					{$^{154}$Dy}  &{\scriptsize Exp}&2.23&2.0&3.1&&&&&&\\
					&{\scriptsize D}&2.18&2.58&2.18&0.04&1.33&0.15&-&0.45&\\
					&{\scriptsize K}&2.33&3.91&2.33&0.001&-&-&73&1.15&\\
					&{\scriptsize IFSW}&2.19&3.03&2.2&0.01&-&-&-&1.36&\\
					{$^{186}$Pt} &{\scriptsize Exp}&2.56&2.5& 3.2&&&&&&\\
					&{\scriptsize D}&2.38&3.46&2.3&0.0020&56.34&0.66&-&0.88&\\
					&	{\scriptsize K}&2.39&5.14&2.3&0.0020&-&-&124&0.96& \\
					&{\scriptsize IFSW}&2.23&3.11&2.2& 0.1999&-&-&-&0.50& \\
					{$^{188}$Pt} &{\scriptsize Exp}&2.53&3.0& 2.3&&&&&&\\
					&{\scriptsize D}&2.34&3.21&2.3&0.0014&39.94&0.53&-&0.62&\\
					&	{\scriptsize K}&2.37&4.59&2.3&0.0040&-&-&100&0.85& \\
					&{\scriptsize IFSW}&2.26&3.18&2.2&0.0029&-&-&-&0.67 &\\	
					{$^{190}$Pt} &{\scriptsize Exp}&2.49&3.1& 2.0&&&&&&\\
					&{\scriptsize D}&2.27&3.40&2.2&0.0010&5.88&0.86&-&0.74&\\
					&	{\scriptsize K}&2.36&4.42&2.3&0.0038&-&-&93&0.67& \\
					&{\scriptsize IFSW}&2.19&3.03&2.2&0.0010&-&-&-&0.75&  \\
					{$^{192}$Pt} &{\scriptsize Exp}&2.48&3.8& 1.9&&&&&&\\
					&{\scriptsize D}&2.29&3.62&2.2&0.0050&1.65&1.04&-&0.52&\\
					&	{\scriptsize K}&2.35&4.31&2.3&0.0020&-&-&88&0.65 &\\
					&{\scriptsize IFSW}&2.20&3.04&2.2&0.0099&-&-&-&0.58  &\\
					{$^{194}$Pt} &{\scriptsize Exp}&2.47&3.9& 1.9&&&&&&\\
					&{\scriptsize D}&2.28&3.71&2.2&0.0040&0.81&1.15&-&0.55&\\
					&	{\scriptsize K}&2.35&4.25&2.3&0.0020&-&-&86&0.71&\\
					&{\scriptsize IFSW}&2.22&3.09&2.2&0.0125&-&-&-&0.61 & \\
					{$^{196}$Pt} &{\scriptsize Exp}&2.47&3.2& 1.9&&&&&&\\
					&{\scriptsize D}&2.30&2.80&2.3&0.0015&39.99&0.86&-&0.56&\\
					&	{\scriptsize K}&2.33&4.01&2.3&0.0040&-&-&76&0.75& \\
					&{\scriptsize IFSW}&2.20&3.04&2.2&0.0155&-&-&-&0.61& \\
					{$^{198}$Pt} &{\scriptsize Exp}&2.42&3.2& 1.9&&&&&&\\
					&{\scriptsize D}&2.00&2.37&2.&0.0040&10.18&0&-&0.34&\\
					&	{\scriptsize K}&2.17&2.59&2.1&0.0050&-&-&31&0.37& \\
					&{\scriptsize IFSW}&2.19&3.03&2.2&0.0050&-&-&-&0.72 &\\
					{$^{200}$Pt} &{\scriptsize Exp}&2.35&2.4& 1.8&&&&&&\\
					&{\scriptsize D}&2.00&2.06&2.0&0.0099&0.25&0&-&0.28&\\
					&	{\scriptsize K}&2.00&1.96&2.0&0.0099&-&-&16&0.27& \\
					&{\scriptsize IFSW}&2.19&3.03&2.20&0.0100&-&-&-&0.64 & \\
					\hline 
	\end{tabular}}}}
\end{table*}

\begin{table*}[tbph]
	\centering
	{\renewcommand{\arraystretch}{1}
		{\setlength{\tabcolsep}{0.5cm} 
			\caption{Comparison of experimental data \cite{b54} (upper line) of the $ \gamma $-unstable Bohr Hamiltonien to theoretical predictions with DDM formalism (lower lines) with $ (\delta=\lambda=0) $ for Davidson (D), Kratzer (K) and Infinite Square Well (ISW) potential respectevely.
			}
			\label{Table2}
			\resizebox{0.90\textwidth}{!}{%
				\begin{tabular}{|ccccccccccc|}		
					\hline 
					
					Nucleus &&$ {R_{4,2}}$ & $ {R_{0,2}}$&$ {R_{2,2}}$&a&c&$ \beta_{0} $&$\tilde{B}  $&$\sigma_{r.m.s.}$&\\
					
					\hline
					&&&&&&&&&&\\	
					{$^{98}$Ru} &{\scriptsize Exp}&2.14&2.0&2.2&&&&&&\\
					&	{\scriptsize D}&2.18&1.66&2.1&0.0100&5.00&1.42&-&0.23&\\
					&	{\scriptsize K}&2.34&2&2.3&0.0101&-&-&38&0.81& \\
					&{\scriptsize IFSW}&2.19&304&2.2&0.0100&-&-&-&0.86&\\
					
					{$^{100}$Ru} &{\scriptsize Exp}&2.27&2.1&2.5 &&&&&&\\
					&	{\scriptsize D}&2.39&2.09&2.4&0.0090&85.36&3.60&-&0.37& \\
					&	{\scriptsize K}&2.40&2.1&2.4&0.0094&-&-&64&0.82& \\
					&{\scriptsize IFSW}&2.19&3.04&2.2&0.0100&-&-&-&0.73&\\
					{$^{102}$Ru} &{\scriptsize Exp}&2.33&2.0&2.3 &&&&&&\\
					&	{\scriptsize D}&2.27&1.98&2.2&0.0132&11.85&2.07&-&0.26&\\
					&	{\scriptsize K}&2.35&2&2.3&0.0105&-&-&41&0.38&\\
					&{\scriptsize IFSW}&2.19&3.03&2.2&0.0100&-&-&-&0.46&\\
					{$^{104}$Ru} &{\scriptsize Exp}&2.48&2.8&2.5 &&&&&&\\
					&	{\scriptsize D}&2.37&2.89&2.3&32.18&3.39&0.0040&-&0.34& \\
					&	{\scriptsize K}&2.39&3.10&2.4&0.0046&-&-&60&0.43&\\
					&{\scriptsize IFSW}&2.17&3.18&2.2&0.0010&-&-&-&0.83& \\
					{$^{102}$Pd}&{\scriptsize Exp}&2.29&2.9&2.8&&&&&&\\
					&{\scriptsize D}& 2.33&1.97&2.3&0.0079&32.10&2.70&-&0.39& \\
					&	{\scriptsize K}&2.42&2.4&2.34&0.0072&-&-&81&0.99&\\
					&{\scriptsize IFSW}&2.19 &3.04&2.2&0.0010&-&-&-&0.84 &\\
					{$^{104}$Pd} &{\scriptsize Exp}&2.38 &2.4&2.4 &&&&&&\\
					&{\scriptsize D}&2.26&2.09 &2.2&8.55&1.98&0.0060&-&0.29&\\
					&	{\scriptsize K}&2.35&2.4&2.3&0.0072&-&-&41&0.32 &\\
					&{\scriptsize IFSW}&2.17&3.18 &2.2&0.0130&-&-&-&0.70& \\
					{$^{106}$Pd} &{\scriptsize Exp}&2.40&2.2 &2.2&& &&&&\\
					&{\scriptsize D}&2.38&4.38&2.4&0.0100&13.53&1.86&-&0.37& \\
					&	{\scriptsize K}&2.33&2.3&2.3&0.0082&-&-&36&0.39 &\\
					&{\scriptsize IFSW}&2.19&3.03 &2.2&0.0010&-&-&-&0.73&\\
					{$^{108}$Pd}&{\scriptsize Exp}&2.42&2.4 &2.1 &&&&&&\\
					&{\scriptsize D}&2.34&2.06 &2.3&35.06&35&2.86&-&0.29&\\
					&	{\scriptsize K}&2.38&2.5&2.4&0.0069&-&-&55&0.31&\\
					&{\scriptsize IFSW}&2.19& 3.03&2.2&0.0010&-&-&-&0.86 &\\
					{$^{110}$Pd}&{\scriptsize Exp}&2.46&2.5 &2.2&&&&&&\\
					&{\scriptsize D}&2.40&2.17&2.4&0.007&116.39&3.98&-&0.30& \\
					&	{\scriptsize K}&2.4&2.7&2.4&0.0061&-&-&100&0.37& \\
					&{\scriptsize IFSW}&2.19& 3.03&2.2&0.0010&-&-&-&0.79 &\\
					{$^{112}$Pd}&{\scriptsize Exp}&2.53 &2.6&2.1 &&&&&&\\
					&{\scriptsize D}& 2.30&2.53&2.3&0.0051&10.39&2.36&-&0.39&\\
					&	{\scriptsize K}&2.32&2.6&2.3&0.0058&-&-&33&0.48&\\
					&{\scriptsize IFSW}&2.19&3.04 &2.2&0.0090&-&-&-&0.46&\\
					{$^{114}$Pd} &{\scriptsize Exp}&2.56 &2.6&2.1&&&&& &\\
					&{\scriptsize D}&2.37&2.62&2.3&0.0090&35.94&3.29&-&0.62 &\\
					&	{\scriptsize K}&2.40&2.6&2.4&0.0065&-&-&65&0.77&\\
					&{\scriptsize IFSW}&2.19&3.03 &2.1&0.0090&-&-&-&0.87&\\
					\hline 
	\end{tabular}}}}
\end{table*}
\begin{table*}[tbph]
	\centering
	\setcounter{table}{1} \caption{(continued)}
	{\renewcommand{\arraystretch}{1}
		{\setlength{\tabcolsep}{0.5cm} 
			\label{Table2p}
			\resizebox{0.90\textwidth}{!}{%
				\begin{tabular}{|ccccccccccc|}		
					\hline 
					Nucleus &&$ {R_{4,2}}$ & $ {R_{0,2}}$&$ {R_{2,2}}$&a&c&$ \beta_{0} $&$\tilde{B}  $&$\sigma_{r.m.s.}$&\\
					\hline
					&&&&&&&&&&\\	
					{$^{116}$Pd}&{\scriptsize Exp}&2.58&3.3 &2.2 &&&&&&\\
					&{\scriptsize D}&2.39&3.15 &2.3&0.0020&40.00&3.76&-&0.56& \\
					&	{\scriptsize K}&2.42&3.3&2.4&0.0044&-&-&83&0.63&\\
					&{\scriptsize IFSW}&2.19&3.04&2.2&0.0100&-&-&--&0.80&\\
					{$^{106}$Cd}&{\scriptsize Exp}&2.36&2.8&2.7&&&&&&\\
					&{\scriptsize D}&2.27&2.83&2.3&0.0050&4.63&2.08&-&0.20 &\\
					&	{\scriptsize K}&2.33&2.8&2.3&0.0044&-&-&36&0.17& \\
					&{\scriptsize IFSW}&2.18&3.1&2.2&0.0010&-&-&-&0.36& \\
					{$^{108}$Cd}  &{\scriptsize Exp}&2.38&2.7&2.52&&&&&& \\
					&{\scriptsize D}&2.18&2.7&2.2&0.0010&1.33&1.49&-&0.40& \\
					&	{\scriptsize K}&2.34&2.7&2.3&0.0054&-&-&39&0.90& \\
					&{\scriptsize IFSW}&2.08&3.51&2.1&0.0099&-&-&-&0.56&\\
					{$^{110}$Cd} &{\scriptsize Exp}&2.35&2.2&2.2&&&&&& \\
					&{\scriptsize D}&2.18&1.22&2.2 &0.0158&11.32&1.41&-&0.35&\\
					&	{\scriptsize K}&2.29&1.9&2.3&0.0115&-&-&28&0.34& \\
					&{\scriptsize IFSW}&2.19&3.03&2.2&0.0010&-&-&-&0.76 &\\
					{$^{114}$Cd} &{\scriptsize Exp}&2.30&2.0&2.2&&&&&& \\
					&{\scriptsize D}&2.12&0.9&2.1 &0.0100&10.79&1.09&-&0.40&\\
					&	{\scriptsize K}&2.25&1.7&2.2 &0.0127&-&-&22&0.24&\\
					&{\scriptsize IFSW}&2.19&3.03&2.2&0.0010&-&-&-& 0.92&\\
					{$^{116}$Cd}  &{\scriptsize Exp}&2.38&2.5&2.4&&&&&&\\
					&{\scriptsize D}&2.22&2.37&2.2&0.0010&3.87&0.33&-&0.29  &\\
					&	{\scriptsize K}&2.27&2.8&2.3&0.0028&-&-&25&0.30& \\
					&{\scriptsize IFSW}&2.06&3.54&2.1&0.0010&-&-&-& 0.54& \\
					\hline 
	\end{tabular}}}}
\end{table*}
\begin{table*}[tbph]
	\centering
	\setcounter{table}{1} \caption{(continued)}
	{\renewcommand{\arraystretch}{1}
		{\setlength{\tabcolsep}{0.5cm} 
			\label{Table3a}
			\resizebox{0.88\textwidth}{!}{%
				\begin{tabular}{|cccccccccc|}		
					\hline 
					Nucleus &&$ {R_{4,2}}$ & $ {R_{0,2}}$&$ {R_{2,2}}$&a&c&$ \beta_{0} $&$ \tilde{B} $&$\sigma_{r.m.s.}$\\
					\hline
					&&&&&&&&&\\
					{$^{118}$Cd}  &{\scriptsize Exp}&2.39&2.6&2.6&&&&&\\
					&{\scriptsize D}&2.19&2.63&2.2&0.0050&1.63&1.53&-&0.37  \\
					&	{\scriptsize K}&2.29&2.6&2.3&0.0045&-&-&28&0.31\\
					&{\scriptsize IFSW}&2.19&3.06&2.2&0.0199&-&-&-&0.64 \\
					{$^{120}$Cd}  &{\scriptsize Exp}&2.38&2.7&2.6&&&&&\\
					&{\scriptsize D}&2.20&2.7&2.2&0.0040&1.77&1.61&-&0.36 \\
					&	{\scriptsize K}&2.31&2.7&2.3&0.0045&-&-&32&0.42\\
					&{\scriptsize IFSW}&2.08&3.52&2.1&0.0010&-&-&-& 0.43 \\
					{$^{118}$Xe} &{\scriptsize Exp}&2.40&2.5& 2.8&&&&&\\
					&{\scriptsize D}&2.38&2.4&2.3&0.0069&57.21&3.54&-&0.30\\
					&	{\scriptsize K}&2.41&2.8&2.4&0.0058&-&-&77&0.40 \\
					&{\scriptsize IFSW}&2.07&3.52&2.1&0.0100&-&-&-&0.86  \\
					{$^{120}$Xe} &{\scriptsize Exp}&2.47&2.8& 2.7&&&&&\\
					&{\scriptsize D}&2.42&2.7&2.4&0.0039&108.2&4.85&-&0.42\\
					&	{\scriptsize K}&2.45&3.0&2.4&0.0049&-&-&133&0.70 \\
					&{\scriptsize IFSW}&2.17&3.18&2.2&0.0700&-&-&-&0.76  \\
					{$^{122}$Xe} &{\scriptsize Exp}&2.50&3.5& 2.5&&&&&\\
					&{\scriptsize D}&2.43&3.47&2.4&0.0049&100.75&4.97&-&0.58\\
					&	{\scriptsize K}&2.45&3.5&2.4&0.0040&-&-&131&0.73\\
					&{\scriptsize IFSW}&2.20&2.95&2.20&0.1000&-&-&-&0.75  \\	
					{$^{124}$Xe} &{\scriptsize Exp}&2.48&3.6& 2.4&&&&&\\
					&{\scriptsize D}&2.39&3.48&2.4&0.0049&32.22&3.76&-&0.57\\
					&	{\scriptsize K}&2.43&3.7&2.4&0.0035&-&-&93&0.72 \\
					&{\scriptsize IFSW}&2.18&23.11&2.1&0.0500&-&-&-&0.82  \\	
					{$^{126}$Xe} &{\scriptsize Exp}&2.42&3.4& 2.3&&&&&\\
					&{\scriptsize D}&2.36&2.8&2.3&0.0039&24.50&3.17&-&0.57\\
					&	{\scriptsize K}&2.39&3.4&2.4&0.0035&-&-&60&0.60 \\
					&{\scriptsize IFSW}&2.17&3.18&2.20&0.0100&-&-&-&0.86 \\
					{$^{128}$Xe} &{\scriptsize Exp}&2.33&3.6& 2.2&&&&&\\
					&{\scriptsize D}&2.26&3.23&2.3&0.0000&2.96&2.04&-&0.35\\
					&	{\scriptsize K}&2.31&3.7&2.3&0.0000&-&-&32&0.45 \\
					&{\scriptsize IFSW}&2.19&3.03&2.2&0.0000&-&-&-&0.93  \\
					{$^{130}$Xe} &{\scriptsize Exp}&2.25&3.3& 2.1&&&&&\\
					&{\scriptsize D}&2.21&3.3&2.2&0.0069&0.22&1.63&-&0.17\\
					&	{\scriptsize K}&2.30&3.3&2.3&0.0007&-&-&29&0.47\\
					&{\scriptsize IFSW}&2.19&2.03&2.2&0.0010&-&-&-&0.82  \\
					{$^{132}$Xe}&{\scriptsize Exp}&2.16&2.8&1.9&&&&& \\
					&{\scriptsize D}&2.08&2.66&2.0&0.0000&0.1&0.90&-&0.20\\
					&	{\scriptsize K}&1.99&1.93&2.0&00.0000&-&-&9&0.37 \\
					&{\scriptsize IFSW}&2.19&3.04&2.2&0.0099&-&-&-& 0.46 \\
					{$^{134}$Xe}  &{\scriptsize Exp}&2.04&1.9&1.9&&&&& \\
					&{\scriptsize D}&1.98&1.8&2.0&0.0000&0.012&0.10&-&0.58 \\
					&	{\scriptsize K}&1.87&1.6&1.9&0.0000&-&-&4&0.21 \\
					&{\scriptsize IFSW}&2.04&3.59&2.1&0.0099&-&-&-&0.82 \\
					\hline 
	\end{tabular}}}}
\end{table*}
\begin{table*}[tbph]
	\centering
	\setcounter{table}{1} \caption{(continued)}
	{\renewcommand{\arraystretch}{1}
		{\setlength{\tabcolsep}{0.5cm} 
			\label{Table3a}
			\resizebox{0.88\textwidth}{!}{%
				\begin{tabular}{|cccccccccc|}		
					\hline 
					Nucleus &&$ {R_{4,2}}$ & $ {R_{0,2}}$&$ {R_{2,2}}$&a&c&$ \beta_{0} $&$ \tilde{B} $&$\sigma_{r.m.s.}$\\
					\hline
					&&&&&&&&&\\
					{$^{130}$Ba} &{\scriptsize Exp}&2.52&3.3&2.5&&&&& \\
					&{\scriptsize D}&2.43&3.4&2.4&0.0099&99.9&4.95&-&0.31\\
					&	{\scriptsize K}&2.45&3.3&2.4&0.0043&-&-&140&0.39\\
					&{\scriptsize IFSW}&2.19&3.06&2.2&0.0000&-&-&-&0.66  \\
					{$^{132}$Ba} &{\scriptsize Exp}&2.43&3.2&2.2&& &&&\\
					&{\scriptsize D}&2.35&3.24&2.3&0.0099&13.4&2.92&-&0.40\\
					&	{\scriptsize K}&2.37&3.2&2.4 &0.0037&-&-&50&0.76\\
					&{\scriptsize IFSW}&2.19&3.03&2.2&0.0090&-&-&-& 0.51 \\
					{$^{134}$Ba}  &{\scriptsize Exp}&2.32&2.9&1.9&&&&&\\
					&{\scriptsize D}&2.12&2.91&2.2&0.0009&0.23&1.12&-&0.26\\
					&	{\scriptsize K}&2.20&2.8&2.2&0.0000&-&-&18&0.34\\
					&{\scriptsize IFSW}&2.19&3.03&2.2&0.0009&-&-&-&0.39 \\
					{$^{136}$Ba} &{\scriptsize Exp}&2.28&1.9& 1.9&&&&&\\
					&{\scriptsize D}&1.99&1.87&2.00&0.0010&0.01&0.23&-&0.18\\
					&	{\scriptsize K}&2.00&1.9&2.0&0.0002&-&-&8&0.19 \\
					&{\scriptsize IFSW}&2.17&3.18&2.2&0.0100&-&-&-&0.68  \\
					{$^{142}$Ba} &{\scriptsize Exp}&2.32&4.27& 3.96&&&&&\\
					&{\scriptsize D}&2.39&4.26&2.4&0.0020&18.16&3.68&-&0.52\\
					&	{\scriptsize K}&2.41&4.3&2.4&0.0021&-&-&79&0.59 \\
					&{\scriptsize IFSW}&2.19&3.02&2.2&0.0020&-&-&-&0.96 \\
					{$^{134}$Ce} &{\scriptsize Exp}&2.56&3.7& 2.4&&&&&\\
					&{\scriptsize D}&2.36&3.6&2.3&0.0010&13.46&3.11&-&0.53\\
					&	{\scriptsize K}&2.42&3.9&2.4&0.0030&-&-&88&0.88 \\
					&{\scriptsize IFSW}&2.21&3.06&2.2&0.0050&-&-&-&0.67  \\
					{$^{136}$Ce} &{\scriptsize Exp}&2.38&1.9& 2.0&&&&&\\
					&{\scriptsize D}&2.15&1.94 &2.1&0.0099&2.07&1.26&-&0.40\\
					&	{\scriptsize K}&2.28&1.9&2.3&0.0105&-&-&27&0.54 \\
					&{\scriptsize IFSW}&2.14&3.3&2.1&0.0149&-&-&-&0.61 \\
					{$^{138}$Ce} &{\scriptsize Exp}&2.32&1.9& 1.9&&&&&\\
					&{\scriptsize D}&1.98&1.79&2.0&0.0050&0.01&0.003&-&0.22\\
					&	{\scriptsize K}&2.13&1.9&2.1&0.0083&-&-&13&0.35 \\
					&{\scriptsize IFSW}&1.94&3.72&2.2&0.0089&-&-&-&0.75 \\
					{$^{140}$Nd} &{\scriptsize Exp}&2.33&1.8&1.9&&&&&\\
					&{\scriptsize D}&2.04&1.82&2&0.0070&0.37&0.66&-&0.13\\
					&	{\scriptsize K}&2.09&1.8&2.1&0.0073&-&-&11&0.16 \\
					&{\scriptsize IFSW}&2.19&3.04&2.2&0.0070&-&-&-&0.60  \\
					{$^{148}$Nd} &{\scriptsize Exp}&2.49&3.0& 4.1&&&&&\\
					&{\scriptsize D}&2.40&2.65&2.4&0.0052&80.41&4.06&-&0.69\\
					&	{\scriptsize K}&2.42&3.3&2.4&0.0042&-&-&90&0.71 \\
					&{\scriptsize IFSW}&2.19&3.03&2.2&0.0010&-&-&-&0.91  \\
					{$^{140}$Sm}&{\scriptsize Exp}&2.35&1.9&2.7&&&&& \\
					&{\scriptsize D}&2.33&1.86&2.3&0.0090&39.85&2.76&-&0.13 \\
					&	{\scriptsize K}&2.36&1.9&2.4&0.0115&-&-&44&0.16 \\
					&{\scriptsize IFSW}&2.14&3.31&2.1&0.0100& -&-&-&0.57 \\
					{$^{142}$Sm}  &{\scriptsize Exp}&2.33&1.9&2.2&& &&&\\
					&{\scriptsize D}&2.08&1.88&2.1&0.0080&0.73&0.88&-&0.10 \\
					&	{\scriptsize K}&2.16&1.9&2.2&0.0089&-&-&15&0.11\\
					&{\scriptsize IFSW}&2.19&3.03&2.2&0.0010&-&-&-&0.60  \\
					\hline 
	\end{tabular}}}}
\end{table*}
\begin{table*}[tbph]
	\centering
	\setcounter{table}{1} \caption{(continued)}
	{\renewcommand{\arraystretch}{0.95}
		{\setlength{\tabcolsep}{0.5cm} 
			\label{Table3aa}
			\resizebox{0.85\textwidth}{!}{%
				\begin{tabular}{|cccccccccc|}		
					\hline 
					Nucleus &&$ {R_{4,2}}$ & $ {R_{0,2}}$&$ {R_{2,2}}$&a&c&$ \beta_{0} $&$\tilde{B}  $&$\sigma_{r.m.s.}$\\
					\hline
					&&&&&&&&&\\
					{$^{142}$Gd} &{\scriptsize Exp}&2.35&2.7&1.9&&&&& \\
					&{\scriptsize D}&2.23&2.65&2.2 &0.0050&3.06&1.79&-&0.18\\
					&	{\scriptsize K}&2.33&2.6&2.3&0.0054&-&-&35&0.29 \\
					&{\scriptsize IFSW}&2.17&3.18&2.2&0.0099&-&-&-&0.39 \\
					{$^{144}$Gd} &{\scriptsize Exp}&2.35&2.5&2.5&&&&& \\
					&{\scriptsize D}&2.33&2.53&2.5&0.0060&19.16&2.76&-&0.08\\
					&	{\scriptsize K}&2.35&2.5&2.4 &0.0065&-&-&11&0.10\\
					&{\scriptsize IFSW}&2.19&3.03&2.2&0.0050&-&-&-& 0.30\\
					{$^{152}$Gd}  &{\scriptsize Exp}&2.19&1.8&3.2&&&&&\\
					&{\scriptsize D}&2.26&0.81&2.2 &0.02&100&2.05&-&0.47 \\
					&	{\scriptsize K}&2.34&1.9&2.4 &0.0116&-&-&40&0.38 \\
					&{\scriptsize IFSW}&2.16&2.69&2.16&0.0010&-&-&-&0.80  \\
					{$^{154}$Dy}  &{\scriptsize Exp}&2.23&2.0&3.1&&&&&\\
					&{\scriptsize D}&2.28&2.67&2.28&0.06&4.09&1.95&-&0.33\\
					&	{\scriptsize K}&2.40&1.7&2.4&0.012&-&-&67&0.94\\
					&	{\scriptsize IFSW}&1.96&1.71&2.0&0.01&-&-&-&1.47\\
					{$^{186}$Pt} &{\scriptsize Exp}&2.56&2.5& 3.2&&&&&\\
					&{\scriptsize D}&2.46&3.7&2.4&0.002&397&6.78&-&0.72\\
					&	{\scriptsize K}&2.47&3.6&2.5&0.0035&-&-&249&0.79\\
					&{\scriptsize IFSW}&2.19&3.03&2.2&0.0010&-&-&-&0.90  \\
					{$^{188}$Pt} &{\scriptsize Exp}&2.53&3.0& 2.3&&&&&\\
					&{\scriptsize D}&2.41&3.0&2.4&0.001&68.95&4.21&-&0.47\\
					&	{\scriptsize K}&2.43&3.2&2.4&0.0047&-&-&100&0.45 \\
					&{\scriptsize IFSW}&2.19&3.04&2.2&0.0010&-&-&-&0.95 \\
					{$^{190}$Pt} &{\scriptsize Exp}&2.49&3.1& 2.0&&&&&\\
					&{\scriptsize D}&2.30&3.5&2.3&0.001&4.53&2.41&-&0.72\\
					&	{\scriptsize K}&2.37&3.2&2.4&0.0038&-&-&49&0.53\\
					&{\scriptsize IFSW}&2.17&3.13&2.2&0.0010&-&-&-&0.65  \\	{$^{192}$Pt} &{\scriptsize Exp}&2.48&3.8& 1.9&&&&&\\
					&{\scriptsize D}&2.30&3.6&2.3&0.0049&2.53&2.33&-&0.54\\
					&	{\scriptsize K}&2.38&3.8&2.4&0.0021&-&-&53&0.69 \\
					&{\scriptsize IFSW}&2.17&3.18&2.20&0.0050&-&-&-&0.76  \\
					{$^{194}$Pt} &{\scriptsize Exp}&2.47&3.9&& 1.9&&&&\\
					&{\scriptsize D}&2.28&4.23&2.3&0.0049&1.87&2.21&-&0.58\\
					&	{\scriptsize K}&2.39&3.9&2.4&0.0023&-&-&60&0.68 \\
					&{\scriptsize IFSW}&2.11&3.42&2.1&0.0199&-&-&-&0.70 \\
					{$^{196}$Pt} &{\scriptsize Exp}&2.47&3.2& 1.9&&&&\\
					&{\scriptsize D}&2.29&3.51&2.3&0.0049&3.87&2.29&-&0.53\\
					&	{\scriptsize K}&2.38&3.1&2.4&0.0043&-&-&54&0.67 \\
					&{\scriptsize IFSW}&2.17&3.18&2.2&0.0050&-&-&-&0.75 \\
					{$^{198}$Pt} &{\scriptsize Exp}&2.42&3.2& 1.9&&&&&\\
					&{\scriptsize D}&2.23&2.6&2.2&0.0049&2.67&1.75&-&0.38\\
					&	{\scriptsize K}&2.25&2.3&2.3&0.0059&-&-&23&0.37 \\
					&{\scriptsize IFSW}&2.19&3.04&2.2&0.0050&-&-&-&0.64 \\
					{$^{200}$Pt} &{\scriptsize Exp}&2.35&2.4& 1.8&&&&&\\
					&{\scriptsize D}&2.02&2.1&2.0&0.0000&0.01&0.52&-&0.31\\
					&	{\scriptsize K}&2.00&1.9&2.0&0.0000&-&-&8&0.34 \\
					&{\scriptsize IFSW}&2.01&3.65&2.0&0.0000&-&-&-&0.58\\
					\hline 
	\end{tabular}}}}
\end{table*}
\begin{table*}
	\centering
	\caption{Comparison of experimental data \cite{b54} (upper line) for several $B(E2)$ ratios of $\gamma$-unstable nuclei
		to predictions (lower line) by the Bohr Hamiltonian with ML formalism  for the Kratzer and Davidson potentials.}
	
	\label{Table3}
	\bigskip
	\resizebox{0.85\textwidth}{!}{%
		\begin{tabular}{l r@{.}l r@{.}l r@{.}l r@{.}l r@{.}l r@{.}l r@{.}l r@{.}l r@{.}l r@{.}l}
			
			\hline
			\multicolumn{1}{l}{nucl.}
			&\multicolumn{2}{l} {}
			&\multicolumn{2}{c} {$4_1\to 2_1 \over 2_1\to 0_1$}
			&\multicolumn{2}{c} {$6_1\to 4_1 \over 2_1\to 0_1$}
			&\multicolumn{2}{c} {$8_1\to 6_1 \over 2_1\to 0_1$}
			&\multicolumn{2}{c} {$10_1\to 8_1 \over 2_1\to 0_1$}
			&\multicolumn{2}{c} {$2_2 \to 2_1 \over 2_1\to 0_1$}
			&\multicolumn{2}{c}{$2_2 \to 0_1 \over 2_1\to 0_1$}
			&\multicolumn{2}{c}{$0_2 \to 2_1 \over 2_1\to 0_1$}
			&\multicolumn{2}{c}{$2_3 \to 0_1 \over 2_1 \to 0_1$}  \\
			
			& \omit\span & \omit\span & \omit\span & \omit\span &
			\omit\span &  \multicolumn{2}{c} {x $10^3$} &  \omit\span &
			\multicolumn{2}{c} {x $10^3$}
			\\

			\hline 
			\\
			$^{98}$Ru &Exp&  & 1&44(25) & \omit\span & \omit\span & \omit\span &
			1&62(61) &  36&0(152)      & \omit\span &
			\omit\span \\
			&	D &	&1&42&1&57&1&65&1&69&1&42&0&0&0&02&3&58&\\
			\omit\span \\
			&	K&	& 1&55 & 1&92 & 2&30 & 2&76 & 1&55 & 0&0 & 0&93 & 0&13 \\
			$^{100}$Ru &Exp&  & 1&45(13) & \omit\span & \omit\span & \omit\span &
			0&64(12) &  41&1(52)      & 0&98(15) & \omit\span \\
			&D&&1&99&2&82&3&61&4&38&1&99&0&0&1&98&0&6&\\
			&	K&	& 1&55 & 1&92 & 2&31 & 2&77 & 1&55 & 0&0 & 0&93 & 0&134 \\
			
			$^{102}$Ru&Exp&   & 1&50(24) & \omit\span & \omit\span & \omit\span &
			0&62(7) &  24&8(7)      & 0&80(14) & \omit\span \\
			&D&&1&43&1&58&1&67&1&73&1&43&0&0&&057&8&29&\\
			&K&	& 1&55 & 1&91 & 2&29 & 2&73 & 1&55 & 0&0 & &92 & 0&132 \\
			
			$^{104}$Ru  &Exp& & 1&18(28) & \omit\span & \omit\span & \omit\span &
			0&63(15) &  35&0(84)      & 0&42(7) & \omit\span \\
			&D&&1&43&1&58&1&67&1&72&1&43&0&0&0&05&7&85&\\
			&K&	& 1&56 & 1&95 & 2&37& 2&87 & 1&56 & 0&0 & 0&98 &0&138 \\
			
			$^{102}$Pd &Exp&  & 1&56(19) & \omit\span & \omit\span & \omit\span &
			0&46(9) &  128&8(735)      & \omit\span & \omit\span \\
			&D&&1&46&1&67&1&82&1&96&1&46&0&0&0&18&14&94&\\
			&K&	& 1&52 & 1&84 & 2&14 & 2&50 & 1&52 & 0&0 & 0&80 & 0&119 \\
			
			$^{104}$Pd &Exp&  & 1&36(27) & \omit\span & \omit\span & \omit\span &
			0&61(8) &  33&3(74)      & \omit\span & \omit\span \\
			&D&	& 1&43 & 1&57 & 1&65 & 1&70 & 1&43 & 0&0 & 0&03& 5&36 \\  
			&K&&1&55&1&91&2&28&2&73&1&55&0&0&0&92&0&132\\
			$^{106}$Pd & Exp& & 1&63(28) & \omit\span & \omit\span & \omit\span &
			0&98(12) &  26&2(31)      & 0&67(18) & \omit\span \\
			&D&	& 1&99 & 2&82 & 3&61 & 4&38 & 1&99 & 0&0 & 1&98 & 0&005 \\
			&K&&1&57&1&98&2&42&2&96&1&57&0&0&1&02&0&142&\\
			$^{108}$Pd  &Exp& & 1&47(20) & 2&16(28) & 2&99(48) & \omit\span &
			1&43(14) &  16&6(18)      & 1&05(13) & 1&90(29) \\
			&D&&1&99&2&83&3&62&4&39&1&99&0&0&1&99&0&0&\\
			&K&	& 1&55 & 1&90 & 2&27 & 2&71 & 1&55 & 0&0 & 0&91 & 0&131 \\ 
			
			$^{110}$Pd&Exp&   & 1&71(34) & \omit\span & \omit\span & \omit\span &
			0&98(24) &  14&1(22)      & 0&64(10) & \omit\span \\
			&D&&1&99&2&83&3&62&4&39&1&99&0&0&1&99&0&0\\
			&K&& 1&54 & 1&89 & 2&25 & 2&67 & 1&54 & 0&0 & 0&89 & 0&129 \\ 
			
			$^{106}$Cd&Exp&   & 1&78(25) & \omit\span & \omit\span & \omit\span &
			0&43(12) &  93&0(127)      & \omit\span & \omit\span \\
			&D&&1&43&1&57&1&65&1&70&1&43&0&0&0&02&4&60\\
			&K&	& 1&57 & 1&97 & 2&41 & 2&95 & 1&57 & 0&0 & 1&02 & 0&141 \\
			
			$^{108}$Cd&Exp&   & 1&54(24) & \omit\span & \omit\span & \omit\span &
			0&64(20) &  67&7(120)      & \omit\span & \omit\span \\
			&D&&1&43&1&59&1&68&1&75&1&43&0&0&0&07&9&62&\\
			&K&	& 1&53 & 1&87 & 2&21 & 2&60& 1&53 & 0&0 & 0&85 & 0&125 \\
			$^{110}$Cd&Exp&   & 1&68(24) & \omit\span & \omit\span & \omit\span &
			1&09(19) &  48&9(78)      & \omit\span & 9&85(595) \\
			&D&&1&99&2&83&3&62&4&39&1&99&0&0&1&99&0&0\\
			&K&	& 1&62 & 2&10& 2&66 &3&37 & 1&62 & 0&0 & 1&19 & 0&15 \\
			
			$^{112}$Cd&Exp&   & 2&02(22) & \omit\span & \omit\span & \omit\span &
			0&50(10) &  19&9(35)      & 1&69(48) & 11&26(210) \\
			&D&&1&99&2&83&3&62&4&39&1&99&0&0&1&99&0&0\\
			&K&& 1&68 & 2&27 & 3&01& 3&98 & 1&68 & 0&0 & 1&41 & 0&17 \\ 
			
			$^{114}$Cd &Exp&  & 1&99(25) & 3&83(72) & 2&73(97) & \omit\span &
			0&71(24) &  15&4(29)      & 0&88(11) & 10&61(193) \\
			&D&&1&99&2&83&3&62&4&39&1&99&0&0&1&99&0&0\\
			&	K&	& 1&67& 2&25 & 2&96 & 3&89 & 1&67& 0&0 & 1&38& 0&17 \\   
			
			$^{116}$Cd &Exp&  & 1&70(52) & \omit\span & \omit\span & \omit\span &
			0&63(46) &  32&8(86)      & 0&02 & \omit\span \\
			&D&&1&43&1&57&1&65&1&70&1&43&0&0&0&043&5&6\\
			&K&	& 1&62 & 2&10 & 2&66 & 3&36 & 1&62 & 0&0 & 1&19 & 0&15\\  
			
			$^{118}$Cd&Exp&   & $>$1&85 & \omit\span & \omit\span & \omit\span &
			\omit\span & \omit\span      & 0&16(4) & \omit\span \\
			&D&&1&42&1&57&1&65&1&70&1&42&0&0&0&026&4&45\\
			&K&& 1&60 & 2&04 & 2&55 & 3&18 & 1&60 & 0&0 & 1&12& 0&15\\  
			$^{120}$Cd&Exp&   & $>$1&85 & \omit\span & \omit\span & \omit\span &
			\omit\span & \omit\span      & 0&16(4) & \omit\span \\
			&D&&1&42&1&57&1&65&1&70&1&42&0&0&0&025&4&43\\
			&K&& 1&58 & 1&98 & 2&43 & 2&98& 1&58 & 0&0 & 1&03& 0&14\\ 
			
			\hline
	\end{tabular}}
\end{table*}  
\begin{table*}
	\centering
	\setcounter{table}{2} \caption{ (continued) }
	
	\bigskip
	\resizebox{0.85\textwidth}{!}{%
		\begin{tabular}{l r@{.}l r@{.}l r@{.}l r@{.}l r@{.}l r@{.}l r@{.}l r@{.}l r@{.}l r@{.}l}

			\hline
			\multicolumn{1}{l}{nucl.}
			&\multicolumn{2}{l} {}
			&\multicolumn{2}{c} {$4_1\to 2_1 \over 2_1\to 0_1$}
			&\multicolumn{2}{c} {$6_1\to 4_1 \over 2_1\to 0_1$}
			&\multicolumn{2}{c} {$8_1\to 6_1 \over 2_1\to 0_1$}
			&\multicolumn{2}{c} {$10_1\to 8_1 \over 2_1\to 0_1$}
			&\multicolumn{2}{c} {$2_2 \to 2_1 \over 2_1\to 0_1$}
			&\multicolumn{2}{c}{$2_2 \to 0_1 \over 2_1\to 0_1$}
			&\multicolumn{2}{c}{$0_2 \to 2_1 \over 2_1\to 0_1$}
			&\multicolumn{2}{c}{$2_3 \to 0_1 \over 2_1 \to 0_1$}  \\

			& \omit\span & \omit\span & \omit\span & \omit\span &
			\omit\span &  \multicolumn{2}{c} {x $10^3$} &  \omit\span &
			\multicolumn{2}{c} {x $10^3$}\\
			\hline
			\\
			
			$^{118}$Xe &Exp&  & 1&11(7) & 0&88(27) & 0&49(20) & $>$0&73 &
			\omit\span & \omit\span & \omit\span &\omit\span \\
			&D&&1&99&2&83&3&62&4&39&1&99&0&0&1&99&0&0\\
			&K&	& 1&52& 1&82 &2&12& 2&46& 1&52 & 0&0 & 0&77& 0&11 \\  
			
			$^{120}$Xe&Exp   && 1&16(14) & 1&17(24) & 0&96(22) & 0&91(19) &
			\omit\span & \omit\span & \omit\span &\omit\span \\
			&D&&1&43&1&57&1&65&1&70&1&43&0&0&0&029&4&93\\
			&K&	& 1&52 & 1&82 & 2&11 & 2&45& 1&52& 0&0 & 0&77 & 0&11\\            
			
			$^{122}$Xe &Exp&  & 1&47(38) & 0&89(26) & $>$0&44 & \omit\span &
			\omit\span & \omit\span    & \omit\span & \omit\span \\
			&D&&1&42&1&57&1&65&1&70&1&42&0&0&0&026&4&52\\
			&K&	& 1&50 & 1&11 & 2&01& 2&28& 1&50 & 0&0 & 0&67 & 0&10\\
			
			$^{124}$Xe &Exp&  & 1&34(24) & 1&59(71) & 0&63(29) & 0&29(8) &
			0&70(19) &  15&9(46)      & \omit\span  & \omit\span \\
			&D&&1&42&1&57&1&65&1&70&1&42&0&0&0&024&4&15\\
			&K&	& 1&49 & 1&75& 1&97 & 2&21 & 1&49 & 0&0 & 0&63 & 0&10\\	
			$^{128}$Xe &Exp&  & 1&47(20) & 1&94(26) & 2&39(40) & 2&74(114) &
			1&19(19) &  15&9(23)      & \omit\span  & \omit\span \\
			&D&&1&42&1&57&1&65&1&69&1&42&0&0&0&020&3&74\\
			&k&	& 1&57   & 1&98 & 2&42 & 2&95& 1&57 & 0&0 & 1&02 & 0&14\\
			
			$^{132}$Xe&Exp&   & 1&24(18) & \omit\span & \omit\span & \omit\span &
			1&77(29) &  3&4(7)      & \omit\span & \omit\span \\
			&D&&1&43&1&58&1&66&1&72&1&43&0&0&0&051&7&67\\
			&K&	& 1&75 & 2&50 & 3&48& 4&81 & 1&75 & 0&0 & 1&68 & 0&19 \\  
			
			$^{130}$Ba &Exp&  & 1&36(6) & 1&62(15) & 1&55(56) & 0&93(15) &
			\omit\span  & \omit\span   & \omit\span  & \omit\span \\
			&D&&1&43&1&57&1&65&1&70&1&43&0&0&0&027&4&68\\
			&K&	& 1&50& 1&76& 2&01&2&27& 1&50 & 0&0 & 0&66 & 0&10\\
			
			$^{132}$Ba &Exp&  & \omit\span & \omit\span & \omit\span & \omit\span &
			3&35(64) &  90&7(177)      & \omit\span & \omit\span \\
			&D&&1&43&1&57&1&65&1&70&1&43&0&0&0&027&4&68\\
			&K&	& 1&52& 1&84 & 2&15& 2&51 & 1&52 & 0&0 & 0&81 & 0&12 \\  
			
			$^{134}$Ba &Exp&  & 1&55(21) & \omit\span & \omit\span & \omit\span &
			2&17(69) &  12&5(41)      & \omit\span & \omit\span \\
			&D&&1&43&1&58&1&65&1&71&1&43&0&0&0&04&6&39\\
			&K&	& 1&68 & 2&28 & 3&03 & 4&01 & 1&68 & 0&0 & 1&42 & 0&17\\    
			
			$^{142}$Ba &Exp&  & 1&40(17) & 0&56(14) & \omit\span & \omit\span &
			\omit\span & \omit\span   & \omit\span & \omit\span \\
			&D&&1&42&1&57&1&64&1&69&1&42&0&0&0&014&2&69\\
			&K&	& 1&95 & 3&09 & 4&74 & 7&14 & 1&95 & 0&0 & 2&28 & 0&22\\ 
			$^{148}$Nd&Exp&   & 1&61(13) & 1&76(19) & \omit\span & \omit\span &
			0&25(4) & 9&3(17)   & 0&54(6) & 32&82(816) \\
			&D&&1&43&1&58&1&66&1&72&1&43&0&0&0&049&7&44\\
			&K&& 1&54 & 1&87 & 2&21 & 2&61 & 1&54& 0&0 & 0&85 & 0&12\\ 
			$^{152}$Gd& Exp & & 1&84(29) & 2&74(81) & \omit\span & \omit\span &
			0&23(4) & 4&2(8)   & 2&47(78) & \omit\span \\
			&D&&1&99&2&83&3&62&4&39&1&99&0&0&1&99&0&0\\
			&K&	& 1&58 & 2&00& 2&47& 3&04& 1&58 & 0&0 & 1&00 & 0&14 \\
			
			$^{154}$Dy  &Exp && 1&62(35) & 2&05(42) & 2&27(62) & 1&86(69) &
			\omit\span   & \omit\span & \omit\span & \omit\span \\
			&D	&& 1&43 & 1&58 & 1&67 & 1&72 & 1&43 & 0&0 & 0&05 & 7&77 \\
			&K&&1&56&1&93&2&32&2&79&1&56&0&0&0&94&134&81\\
			$^{192}$Pt &Exp & & 1&56(12) & 1&23(55) & \omit\span & \omit\span &
			1&91(16)   & 9&5(9) & \omit\span & \omit\span \\
			&D&&1&42&1&57&1&65&1&69&1&42&0&0&0&020&3&99\\
			&K&	& 1&53 & 1&87 & 2&2 & 2&59 & 1&53 & 0&0 & 0&85 & 0&12 \\
			
			$^{194}$Pt &Exp&  & 1&73(13) & 1&36(45) & 1&02(30) & 0&69(19) &
			1&81(25) &  5&9(9)      & 0&01  & \omit\span \\
			&D&&1&42&1&57&1&65&1&69&1&42&0&0&0&02&3&76\\
			&K&	& 1&54& 1&87& 2&22 & 2&61& 1&54 & 0&0 & 0&86 & 0&12\\
			
			$^{196}$Pt &Exp&  & 1&48(3) & 1&80(23) & 1&92(23) & \omit\span &
			\omit\span &  0&4      & 0&07(4)  & 0&06(6) \\
			&D&&1&43&1&58&1&66&1&71&1&43&0&0&0&045&6&98\\
			&K&	& 1&55 & 1&91 & 2&28 & 2&73 & 1&55 & 0&0 & 0&92 & 0&13\\ 
			
			$^{198}$Pt&Exp&   & 1&19(13) & $>$1&78 & \omit\span & \omit\span &
			1&16(23) &  1&2(4)      & 0&81(22)  & 1&56(126) \\
			&D&&1&99&2&83&3&62&4&39&1&99&0&0&1&99&0&0\\
			&K&	& 1&55 & 1&91 & 2&28 & 2&73 & 1&55 & 0&0 & 0&92& 0&13\\   
			\hline 
			\hline
		\end{tabular}
	}
\end{table*} 
\begin{table*}
	\centering
	\caption{Comparison of experimental data \cite{b54} (upper line) for several $B(E2)$ ratios of $\gamma$-unstable nuclei
		to predictions (lower line) by the Bohr Hamiltonian with $\beta$-dependent mass for the Kratzer and Davidson potentials, for the parameter values shown in Table I.}
	\bigskip
	\label{Table4}
	\resizebox{0.85\textwidth}{!}{%
		\begin{tabular}{l r@{.}l r@{.}l r@{.}l r@{.}l r@{.}l r@{.}l r@{.}l r@{.}l r@{.}l r@{.}l}

			\hline
			\multicolumn{1}{l}{nucl.}
			&\multicolumn{2}{l} {}
			&\multicolumn{2}{c} {$4_1\to 2_1 \over 2_1\to 0_1$}
			&\multicolumn{2}{c} {$6_1\to 4_1 \over 2_1\to 0_1$}
			&\multicolumn{2}{c} {$8_1\to 6_1 \over 2_1\to 0_1$}
			&\multicolumn{2}{c} {$10_1\to 8_1 \over 2_1\to 0_1$}
			&\multicolumn{2}{c} {$2_2 \to 2_1 \over 2_1\to 0_1$}
			&\multicolumn{2}{c}{$2_2 \to 0_1 \over 2_1\to 0_1$}
			&\multicolumn{2}{c}{$0_2 \to 2_1 \over 2_1\to 0_1$}
			&\multicolumn{2}{c}{$2_3 \to 0_1 \over 2_1 \to 0_1$}  \\
			
			& \omit\span & \omit\span & \omit\span & \omit\span &
			\omit\span &  \multicolumn{2}{c} {x $10^3$} &  \omit\span &
			\multicolumn{2}{c} {x $10^3$}
			\\

			\hline 
			\\
			$^{98}$Ru &Exp&  & 1&44(25) & \omit\span & \omit\span & \omit\span &
			1&62(61) &  36&0(152)      & \omit\span &
			\omit\span \\
			&	D &	&1&72&2&26&2&79&3&32&1&72&0&0&1&02&5&8&\\
			\omit\span \\
			&	K&	& 1&77 & 2&81 & 4&63 & 8&42 & 1&77 & 0&0 & 1&27 & 27&84 \\
			
			$^{100}$Ru &Exp&  & 1&45(13) & \omit\span & \omit\span & \omit\span &
			0&64(12) &  41&1(52)      & 0&98(15) & \omit\span \\
			&D&&1&56&1&83&1&99&2&09&1&56&0&0&0&99&86&82&\\
			&	K&	& 1&70 & 2&56 & 3&93 & 6&59 & 1&70 & 0&0 & 1&11 & 43&07 \\
			
			$^{102}$Ru&Exp&   & 1&50(24) & \omit\span & \omit\span & \omit\span &
			0&62(7) &  24&8(7)      & 0&80(14) & \omit\span \\
			&D&&1&62&2&05&2&46&2&87&1&62&0&0&0&70&10&40&\\
			&K&	& 1&77 & 2&82 & 4&68 & 8&67 & 1&77 & 0&0 & 1&29 & 31&06 \\
			
			$^{104}$Ru  &Exp& & 1&18(28) & \omit\span & \omit\span & \omit\span &
			0&63(15) &  35&0(84)      & 0&42(7) & \omit\span \\
			&D&&1&63&2&08&2&50&2&93&1&63&0&0&0&74&9&6&\\
			&K&	& 1&60 & 2&20 & 2&95 & 4&00 & 1&60 & 0&0 & 0&68 &25&59 \\
			
			$^{102}$Pd &Exp&  & 1&56(19) & \omit\span & \omit\span & \omit\span &
			0&46(9) &  128&8(735)      & \omit\span & \omit\span \\
			&D&&1&58&1&95&2&30&2&65&1&58&0&0&0&57&13&79&\\
			&K&	& 1&63 & 2&31 & 3&25 & 4&77 & 1&63 & 0&0 & 0&87 & 41&64 \\
			$^{104}$Pd   &Exp&  & 1&36(27) & \omit\span & \omit\span & \omit\span &
			0&61(8) &  33&3(74)      & \omit\span & \omit\span \\
			&D&&1&63&2&08&2&50&2&93&1&63&0&0&0&74&9&69\\
			&K&&1&70&2&52&3&74&5&83&1&70&0&0&0&99&24&16\\
			$^{106}$Pd & Exp& & 1&63(28) & \omit\span & \omit\span & \omit\span &
			0&98(12) &  26&2(31)      & 0&67(18) & \omit\span \\
			&D&&1&65&2&11&2&55&3&0&1&65&0&0&0&79&9&0\\	
			&K&&1&74&2&66&4&13&6&83&1&74&0&0&1&12&22&91\\	
			$^{108}$Pd  &Exp& & 1&47(20) & 2&16(28) & 2&99(48) & \omit\span &
			1&43(14) &  16&6(18)      & 1&05(13) & 1&90(29) \\
			&D&&1&58&1&96&2&31&2&67&1&58&0&0&0&59&15&48&\\
			&K&	& 1&66 & 2&38 & 3&42 & 5&11 & 1&66 & 0&0 & 0&89 & 30&31 \\ 
			
			$^{110}$Pd&Exp&   & 1&71(34) & \omit\span & \omit\span & \omit\span &
			0&98(24) &  14&1(22)      & 0&64(10) & \omit\span \\
			&D&&1&51&1&80&2&04&2&28&1&51&0&0&0&36&16&97&\\
			&K&& 1&60 & 2&18 & 2&94 & 4&06 & 1&60 & 0&0 & 0&75 & 42&18 \\ 
			
			$^{106}$Cd  &Exp& & 1&78(25) & \omit\span & \omit\span & \omit\span &
			0&43(12) &  93&0(127)      & \omit\span & \omit\span \\
			&D&&1&62&2&05&0&45&2&86&1&62&0&0&0&69&10&48\\
			&K&	& 1&66 & 2&37 & 3&34 & 4&76 & 1&66 & 0&0 & 0&83 & 16&97 \\
			
			$^{108}$Cd&Exp&   & 1&54(24) & \omit\span & \omit\span & \omit\span &
			0&64(20) &  67&7(120)      & \omit\span & \omit\span \\
			&D&&1&78&2&34&2&78&3&14&1&78&0&0&1&41&25&45&\\
			&K&	& 1&67 & 2&40 & 3&43 & 5&01 & 1&67 & 0&0 & 0&87 & 19&88 \\
			$^{110}$Cd&Exp&   & 1&68(24) & \omit\span & \omit\span & \omit\span &
			1&09(19) &  48&9(78)      & \omit\span & 9&85(595) \\
			&D&&1&74&2&32&2&87&3&42&1&74&0&0&1&11&5&56&\\
			&K&	& 1&85 & 3&14 & 5&63 &11&54 & 1&85 & 0&0 & 1&52 & 20&99 \\
			
			$^{112}$Cd&Exp&   & 2&02(22) & \omit\span & \omit\span & \omit\span &
			0&50(10) &  19&9(35)      & 1&69(48) & 11&26(210) \\
			&D&&1&78&2&40&3&00&3&59&1&78&0&0&1&24&3&62\\
			&K&& 1&95 & 3&53 & 6&92 & 15&92 & 1&95 & 0&0 & 1&82 & 12&87 \\ 
			
			
			
			$^{118}$Cd &Exp&  & $>$1&85 & \omit\span & \omit\span & \omit\span &
			\omit\span & \omit\span      & 0&16(4) & \omit\span \\
			&D&&1&70&2&22&2&73&2&23&1&70&0&0&0&96&6&63\\
			&K&	& 1&70 & 2&51 & 3&65 & 5&41 & 1&70 & 0&0 & 0&95 & 13&14 \\  
			
			$^{118}$Xe &Exp&  & 1&11(7) & 0&88(27) & 0&49(20) & $>$0&73 &
			\omit\span & \omit\span & \omit\span &\omit\span \\
			&D&&1&58&1&96&2&30&2&64&1&58&0&0&0&62&21&01&\\
			&K&	& 1&61 & 2&21 & 3&00 & 4&17 & 1&61 & 0&0 & 0&74 & 34&42 \\  
			
			$^{120}$Xe&Exp  & & 1&16(14) & 1&17(24) & 0&96(22) & 0&91(19) &
			\omit\span & \omit\span & \omit\span &\omit\span \\
			&D&&1&51&1&79&2&03&2&26&1&51&0&0&0&36&18&22&\\
			&K&	& 1&56 & 2&06 & 2&64 & 3&43 & 1&56 & 0&0 & 0&62 & 42&25 \\            
			
			$^{122}$Xe &Exp&  & 1&47(38) & 0&89(26) & $>$0&44 & \omit\span &
			\omit\span & \omit\span    & \omit\span & \omit\span \\
			&D&&1&52&1&77&1&93&2&04&1&52&0&0&0&76&79&\\
			&K&	& 1&54 & 2&00 & 2&52 & 3&17 & 1&54 & 0&0 & 0&54 & 36&27 \\
			
			$^{124}$Xe &Exp&  & 1&34(24) & 1&59(71) & 0&63(29) & 0&29(8) &
			0&70(19) &  15&9(46)      & \omit\span  & \omit\span \\
			&D&&1&58&1&94&2&28&2&60&1&58&0&0&0&60&22&\\
			&K&	& 1&55 & 2&03 & 2&57 & 3&25 & 1&55 & 0&0 & 0&53 & 28&40 \\
			
			\hline
	\end{tabular}}
\end{table*}  
\begin{table*}
	\centering
	\setcounter{table}{3} \caption{ (continued) }
	
	\bigskip
	\resizebox{0.90\textwidth}{!}{%
		\begin{tabular}{l r@{.}l r@{.}l r@{.}l r@{.}l r@{.}l r@{.}l r@{.}l r@{.}l r@{.}l r@{.}l}

			\hline
			\multicolumn{1}{l}{nucl.}
			&	\multicolumn{2}{c}{}
			&\multicolumn{2}{c} {$4_1\to 2_1 \over 2_1\to 0_1$}
			&\multicolumn{2}{c} {$6_1\to 4_1 \over 2_1\to 0_1$}
			&\multicolumn{2}{c} {$8_1\to 6_1 \over 2_1\to 0_1$}
			&\multicolumn{2}{c} {$10_1\to 8_1 \over 2_1\to 0_1$}
			&\multicolumn{2}{c} {$2_2 \to 2_1 \over 2_1\to 0_1$}
			&\multicolumn{2}{c}{$2_2 \to 0_1 \over 2_1\to 0_1$}
			&\multicolumn{2}{c}{$0_2 \to 2_1 \over 2_1\to 0_1$}
			&\multicolumn{2}{c}{$2_3 \to 0_1 \over 2_1 \to 0_1$}  \\

			& \omit\span & \omit\span & \omit\span & \omit\span &
			\omit\span &  \multicolumn{2}{c} {x $10^3$} &  \omit\span &
			\multicolumn{2}{c} {x $10^3$}
			\\
			\hline 
			\\
			
			$^{128}$Xe &Exp&  & 1&47(20) & 1&94(26) & 2&39(40) & 2&74(114) &
			1&19(19) &  15&9(23)      & \omit\span  & \omit\span \\
			&D&&1&64&2&09&2&53&2&96&1&64&0&0&0&77&10&79&\\
			&K&	& 1&83   & 2&95 & 4&73 & 7&64 & 1&83 & 0&0 & 0&75 & 12&57 \\
			
			$^{132}$Xe&Exp&   & 1&24(18) & \omit\span & \omit\span & \omit\span &
			1&77(29) &  3&4(7)      & \omit\span & \omit\span \\
			&D&&1&61&1&91&2&08&2&19&1&61&0&0&1&17&81&82&\\
			&K&	& 2&78 & 7&13 & 17&89 & 43&35 & 2&78 & 0&0 & 2&49 & 0&07 \\  
			
			$^{130}$Ba &Exp&  & 1&36(6) & 1&62(15) & 1&55(56) & 0&93(15) &
			\omit\span  & \omit\span   & \omit\span  & \omit\span \\
			&D&&1&48&1&69&1&81&1&88&1&48&0&0&0&69&93&53&\\
			&K&	& 1&54 & 2&01 & 2&54 & 3&22 & 1&54 & 0&0 & 0&56 & 39&43 \\
			
			$^{132}$Ba &Exp&  & \omit\span & \omit\span & \omit\span & \omit\span &
			3&35(64) &  90&7(177)      & \omit\span & \omit\span \\
			&D&&1&52&1&74&1&86&1&94&1&52&0&0&0&88&99&49&\\
			&K&	& 1&61 & 2&20 & 2&94 & 3&95 & 1&61 & 0&0 & 0&66 & 20&59 \\ 
			$^{134}$Ba &Exp&  & 1&55(21) & \omit\span & \omit\span & \omit\span &
			2&17(69) &  12&5(41)      & \omit\span & \omit\span \\
			&D&&1&52&1&87&2&03&2&13&1&59&0&0&1&1&85&96&\\
			&K&	& 2&13 & 4&10 & 7&88 & 15&19 & 2&13 & 0&0 & 1&26 & 6&22 \\    
			
			$^{142}$Ba &Exp&  & 1&40(17) & 0&56(14) & \omit\span & \omit\span &
			\omit\span & \omit\span   & \omit\span & \omit\span \\
			&D&&1&97&2&77&3&54&4&29&1&97&0&0&1&89&0&15&\\
			&K&	& 1&54 & 1&99 & 2&46 & 3&04 & 1&54 & 0&0 & 0&45 & 21&34 \\ 
			$^{148}$Nd  &Exp& & 1&61(13) & 1&76(19) & \omit\span & \omit\span &
			0&25(4) & 9&3(17)   & 0&54(6) & 32&82(816) \\
			&D&&1&51&1&80&2&05&10&73&1&78&0&0&1&46&58&09\\
			&K&	& 1&57 & 2&08 & 2&67 & 3&47 & 1&57 & 0&0 & 0&59 & 30&88 \\ 
			$^{152}$Gd &Exp&  & 1&84(29) & 2&74(81) & \omit\span & \omit\span &
			0&23(4) & 4&2(8)   & 2&47(78) & \omit\span \\
			&D&& 1&63&2&08&2&51&2&94&1&63&0&0&0&75&9&58\\
			&K&	& 1&80 & 2&96 & 5&14 & 10&30 & 1&80 & 0&0 & 1&41 & 32&70 \\
			
			$^{154}$Dy &Exp&  & 1&62(35) & 2&05(42) & 2&27(62) & 1&86(69) &
			\omit\span   & \omit\span & \omit\span & \omit\span \\
			&D&&1&63&2&08&2&51&2&93&1&63&0&0&0&77&13&40\\
			&K&	& 1&78 & 2&89 & 5&06 & 10&73 & 1&78 & 0&0 & 1&46 & 58&09 \\
			
			
			$^{192}$Pt&Exp&   & 1&56(12) & 1&23(55) & \omit\span & \omit\span &
			1&91(16)   & 9&5(9) & \omit\span & \omit\span \\
			&D&&1&48&1&67&1&78&1&84&0&0&0&77&106&4&\\
			&K&	& 1&57 & 2&09 & 2&68 & 3&44 & 1&57 & 0&0 & 0&54 & 17&79 \\
			
			$^{194}$Pt &Exp&  & 1&73(13) & 1&36(45) & 1&02(30) & 0&69(19) &
			1&81(25) &  5&9(9)      & 0&01  & \omit\span \\
			&D&&1&48&1&68&1&78&1&85&1&48&0&0&0&77&106&14&\\
			&K&	& 1&56 & 2&07 & 2&63 & 3&34 & 1&56 & 0&0 & 0&52 & 19&45 \\
			
			$^{196}$Pt &Exp&  & 1&48(3) & 1&80(23) & 1&92(23) & \omit\span &
			\omit\span &  0&4      & 0&07(4)  & 0&06(6) \\
			&D&&1&51&1&72&1&85&1&92&1&51&0&0&0&85&101&04&\\
			&K&	& 1&61 & 2&21 & 2&97 & 4&04 & 1&61 & 0&0 & 0&69 & 23&11 \\ 
			
			$^{198}$Pt&Exp&   & 1&19(13) & $>$1&78 & \omit\span & \omit\span &
			1&16(23) &  1&2(4)      & 0&81(22)  & 1&56(126) \\
			&D&&1&57&1&84&1&99&2&09&1&57&0&0&1&04&88&96\\
			&K&	& 1&76 & 2&73 & 4&24 & 6&76 & 1&76 & 0&0 & 1&16 & 11&09 \\          
			
			\hline
	\end{tabular}}
\end{table*}         

\clearpage

\end{document}